\documentclass[twocolumn,english]{aastex6}
\usepackage[T1]{fontenc}
\usepackage[latin9]{inputenc}
\usepackage{units}
\usepackage{bm}
\usepackage{amstext}
\usepackage{amssymb}
\usepackage{graphicx}

\makeatletter
\usepackage{amsmath}
\usepackage{gensymb}

\hypersetup{allcolors=blue}

\journalinfo{The Astronomical Journal, in press (accepted 2019 March 25)}

\makeatother

\usepackage{babel}
\begin{document}

\title{Orbital Deflection of Comets by Directed Energy}

\shorttitle{Comet Deflection: Directed Energy}

\author{Qicheng Zhang\altaffilmark{1,2}}

\author{Philip M. Lubin\altaffilmark{1}}

\author{Gary B. Hughes\altaffilmark{3}}

\email{qicheng@cometary.org}

\affil{$^{1}$ Department of Physics, University of California, Santa Barbara,
CA 93106, USA\\
$^{2}$ Division of Geological and Planetary Sciences, California
Institute of Technology, Pasadena, CA 91125, USA\\
$^{3}$ Statistics Department, California Polytechnic State University,
San Luis Obispo, CA 93407, USA}
\begin{abstract}
Cometary impacts pose a long-term hazard to life on Earth. Impact
mitigation techniques have been studied extensively, but they tend
to focus on asteroid diversion. Typical asteroid interdiction schemes
involve spacecraft physically intercepting the target, a task feasible
only for targets identified decades in advance and in a narrow range
of orbits\textemdash criteria unlikely to be satisfied by a threatening
comet. Comets, however, are naturally perturbed from purely gravitational
trajectories through solar heating of their surfaces which activates
sublimation-driven jets. Artificial heating of a comet, such as by
a laser, may supplement natural heating by the Sun to purposefully
manipulate its path and thereby avoid an impact. Deflection effectiveness
depends on the comet's heating response, which varies dramatically
depending on factors including nucleus size, orbit and dynamical history.
These factors are incorporated into a numerical orbital model to assess
the effectiveness and feasibility of using high-powered laser arrays
in Earth orbit and on the ground for comet deflection. Simulation
results suggest that a diffraction-limited 500~m orbital or terrestrial
laser array operating at 10~GW for 1\% of each day over 1~yr is
sufficient to fully avert the impact of a typical 500~m diameter
comet with primary nongravitational parameter $A_{1}=2\times10^{-8}$~au~d$^{-2}$.
Strategies to avoid comet fragmentation during deflection are also
discussed.
\end{abstract}


\section{Introduction}

Earth-crossing asteroids and comets pose a long-term hazard to human
interests on Earth. Numerous methods to mitigate the impact threat
have been developed, but these generally focus on the asteroid threat
while directing minimal attention toward cometary impactors. These
methods include, but are not limited to:
\begin{enumerate}
\item Kinetic impactors, by which momentum is transferred to the asteroid
via the hypervelocity impact of an expendable spacecraft, optionally
enhanced by an explosive charge \citep{mcinnes:2004,koenig:2007}.
\item Direct application of thrust, via thrusters placed directly onto the
surface of the asteroid \citep{walker:2005} or on one or more gravitationally-bound
spacecraft positioned nearby as ``gravity tractors'' \citep{lu:2005,foster:2013}.
\item Surface albedo alteration, such as by paint \citep{hyland:2010} or
mirrors \citep{vasile:2010} to slowly shift the asteroid's orbit
via radiation pressure.
\end{enumerate}
These strategies all share one fundamental requirement: a spacecraft
must physically intercept the target. This requirement is acceptable
when the target follows a typical low-eccentricity, low-inclination
orbit and is identified decades in advance of a potential impact,
qualities shared by most near-Earth asteroids (NEAs).

These favorable qualities, however, are not common among comets, which
are found at all inclinations and near-parabolic eccentricities. Long-period
comets (LPCs), in particular, are rarely discovered more than 3~yr
in advance of their closest approach to Earth \citep{francis:2005}.
One recent example of an LPC, C/2013 A1 (Siding Spring), approached
to within $140500$~km (0.37~LD) of Mars only 22~months after its
discovery \citep{farnocchia:2016}. While further advancement in sky
survey technology will somewhat improve the warning, discoveries of
these comets remain fundamentally limited by their approach from the
distant outer solar system, where they cannot be observed with telescopes
of any realistic scale. Such short notice leaves little time to even
design an interception mission, much less actually reach the comet
in time to deflect it under a realistic delta-v budget. Any reliable
method for diverting a comet from impact must be capable of operating
remotely and commencing immediately following the identification of
a threat.

One potentially viable approach for deflecting comets is directed
energy heating, whereby a laser array concentrates energy onto the
surface of the target, vaporizing it. The resulting ejecta plume exerts
thrust on the object, shifting it from its original collisional trajectory
\citep{lubin:2014:oe}. One proposed option for deflecting NEAs is,
in fact, to install the laser aboard a rendezvous spacecraft that
intercepts and travels alongside the asteroid. Physical proximity
of the laser, however, is not fundamentally required by the directed
energy approach. The long-range nature of light implies that a laser
array may also be built to operate from Earth orbit, or even from
the ground, deflecting the target remotely. Such a system would permit
an immediate response to any confirmed threat, including an LPC. Directed
energy is particularly applicable as a method for comet deflection
due to the volatile material\textemdash particularly water ice\textemdash on
or near the surface of comets that drive their cometary activity.
Nongravitational accelerations in response to solar heating have been
astrometrically determined for numerous comets \citep{krolikowska:2004}.
The response of these comets to laser heating may then be estimated
by scaling the measured solar heating accelerations under a standard
heating model \citep{delsemme:1971,marsden:1973}.

The effectiveness of near-Earth object deflection via directed energy
has been studied previously for several mission configurations, including
the rendezvous case, in \citet{zhang:2016:pasp}. While cometary impactors
are discussed, they are treated at a cursory level using a crude heating
response model based on the one used for asteroids. The present manuscript
serves as an extension to these earlier results and introduces new
orbital simulations developed specifically to simulate comet deflection
based on existing models of cometary nongravitational forces. In addition,
complications specific to comet deflection, such the risk of fragmentation
under heating, are also briefly addressed.

\section{Simulations}

The simulations model the Sun, the Moon, and the eight known major
planets as Newtonian gravitational point sources in the frame of the
solar system barycenter, with their positions given by JPL DE 421
\citep{folkner:2008}. The comet is treated as a test particle under
the influence of the gravitational sources and of the laser, which
is approximated as coincident with the center of the Earth at position
$\bm{x}_{\oplus}$. Numerical integration is performed with the ``s17odr8a''
composition of the Velocity Verlet method \citep{kahan:1997}.

Note that, while the Moon and planets other than Earth will significant
alter the impact threat posed by a real comet, their inclusion in
the presented simulations has a negligible effect on the results,
as only comets following a direct impact trajectory are simulated.
Prior close encounters with gravitational sources may amplify trajectory
differences and thus improve deflection effectiveness. A random threatening
LPC is unlikely to have had a close encounter with another planet
before reaching Earth, so this scenario will not be further considered
in this analysis.

\subsection{Laser}

Comet deflection is performed via heating of the target comet by a
large array of phased laser elements. Laser pointing\textemdash performed
by adjusting the phasing of the individual elements\textemdash is
assumed to be perfect, with the laser beam exactly centered on the
comet nucleus over the deflection period. Such accurate targeting
may be achieved by scanning the laser beam and monitoring the reflection
of the beam from the nucleus. The resulting astrometry of the nucleus
will aid in constraining the trajectory of the comet both initially
and over the course of the deflection process. This approach is similar
to the one taken in \citet{riley:2014}, which proposes to locate
NEAs by monitoring return signals from a scanning laser beam.

Two classes of laser arrays are considered:
\begin{enumerate}
\item Orbital: the laser array is supported by a photovoltaic (PV) array
operating in low-Earth orbit. Laser output is restricted by both an
operating power $P$, constrained by the number of laser elements
and by their heat dissipation mechanisms, and a time-averaged power
$\langle P\rangle$, constrained by the size and efficiency of the
PV array. In the simulations, the laser operates at $P$ when active;
it is supported by the PV array directly, as well as by an attached
battery system charged by the PV array when the laser is idle. Given
a square PV array of edge length $L_{\text{PV}}$ and total solar-to-laser
efficiency $\varepsilon$, average laser power is $\langle P\rangle=\varepsilon S_{0}L_{\text{PV}}^{2}$.
The simulations consider the laser-carrying spacecraft to be equipped
with a PV array of size $L_{\text{PV}}\sim L_{\text{las}}$, the size
of the laser array itself. While not required in practice, this assumption
is consistent with the orbital laser array designs discussed in \citet{lubin:2014:oe}
in which the PV and laser arrays together comprise the bulk of the
spacecraft's physical size.
\item Terrestrial: the laser array is installed directly on the Earth's
surface. Laser output is restricted to $P$, primarily due to the
size of the array and the number of laser elements available. Electrical
power and heat dissipation capacity impose lesser constraints. Mean
power $\langle P\rangle=f(t)P$ varies over time $t$, where $f(t)$
is the average fraction of time each day the laser can target the
comet. For the laser to be usable, the comet must be within the laser's
field of view of diameter $\Theta_{\text{fov}}$ centered on the zenith.
This condition is dependent on the latitude of the laser $\phi_{\text{las}}$,
as well as on the declination of the comet $\delta_{\text{com}}(t)$
as viewed from Earth. Operation is also constrained by $f(t)\propto\kappa$,
the fraction of acceptable the weather expected at the site of the
laser. Although $\kappa$ may vary on a seasonal basis depending on
local climate, these variations are neglected for the simulations
in which $\kappa$ is considered constant. Careful treatment of $\kappa$
is left to a more detailed study on laser site selection.
\end{enumerate}
Both types of laser arrays are assumed to be capable of producing
a diffraction-limited beam diverging at a half-angle $\theta_{\text{beam}}\approx\lambda_{\text{beam}}/L_{\text{las}}$
for a beam of wavelength $\lambda_{\text{beam}}\approx1$~$\mu$m. 

With a terrestrial array, an adaptive optics system to counteract
atmospheric distortion is necessary to attain such a narrow beam.
Such challenges faced in the construction and operation of these arrays
have been\textemdash and are continuing to be\textemdash analyzed
in detail separately, and will not be discussed in depth here \citep{lubin:2014:oe}.
Unless otherwise noted, the simulations assume these purely engineering
challenges can and will be overcome.

\subsection{Comet}

The target comet is modeled as a non-rotating spherically-symmetric
object with zero thermal inertia, using the semi-empirical nongravitational
acceleration model of \citet{marsden:1973} based on the sublimation
curve of water ice on the comet's surface numerically computed by
\citet{delsemme:1971}. Such a comet at barycentric position $\bm{x}$
illuminated by the Sun at $\bm{x}_{\odot}$, with $r\equiv\|\bm{x}-\bm{x}_{\odot}\|$,
experiences a nongravitational acceleration, produced by jets powered
by sublimating water ice, of

\begin{equation}
\ddot{\bm{x}}_{\text{NG}}=A\times\alpha\left(\frac{r}{r_{0}}\right)^{-m}\left(1+\left(\frac{r}{r_{0}}\right)^{n}\right)^{-k}\frac{\bm{x}-\bm{x}_{\odot}}{r}\label{eq:sun-only}
\end{equation}

with $r_{0}=2.808$~au, $\alpha=0.111262$, $m=2.15$, $n=5.093$,
and $k=4.6142$. This notation is equivalent to $A_{1}=A$, $A_{2}=0$,
$A_{3}=0$ in the original notation of \citet{marsden:1973}, where
$A_{2}$ and $A_{3}$ are analogous to $A$ for the components of
$\ddot{\bm{x}}_{\text{NG}}$ orthogonal to $\bm{x}-\bm{x}_{\odot}$.

Nonzero comet nucleus rotation and thermal inertia will rotate $\ddot{\bm{x}}_{\text{NG}}$
away from the direction of $\bm{x}-\bm{x}_{\odot}$, producing non-radial
components $A_{2}$,~$A_{3}\neq0$. With extremely fast rotation,
$\ddot{\bm{x}}_{\text{NG}}$ weakens in magnitude as the heating thrust
forces become spread over a wide range of directions. More detailed
analyses of this effect are provided by \citet{johansson:2014} and
\citet{griswold:2015}, who discuss the heating response of small,
rotating asteroids. Note that the structural and compositional inhomogeneity
of comets complicates the exact results, although the underlying principles
are similar. The considered $A_{2}=A_{3}=0$ of a non-rotating comet,
however, serves as a good approximation for most comets, which feature
$A_{1}\gg A_{2},A_{3}$ \citep{krolikowska:2004}.

The nongravitational parameter $A$ (the acceleration at $r=1$~au)
has been observationally measured for numerous comets and varies by
several orders of magnitude between different comets depending on
dynamical age, structure, and size \citep{yeomans:2004}. Assuming
thrust $F_{\text{NG}}=m_{\text{com}}\ddot{x}_{\text{NG}}$ is proportional
to the cross sectional area of sunlight intercepted by the comet,
the nongravitational parameter is $A\propto R_{\text{com}}^{-1}\implies A\equiv A_{\unit[1]{km}}\left(\unit[1]{km}/2R_{\text{com}}\right)$
for comets of similar dynamical age and origin of diameter $2R_{\text{com}}$.

\begin{figure*}
\begin{centering}
\plottwo{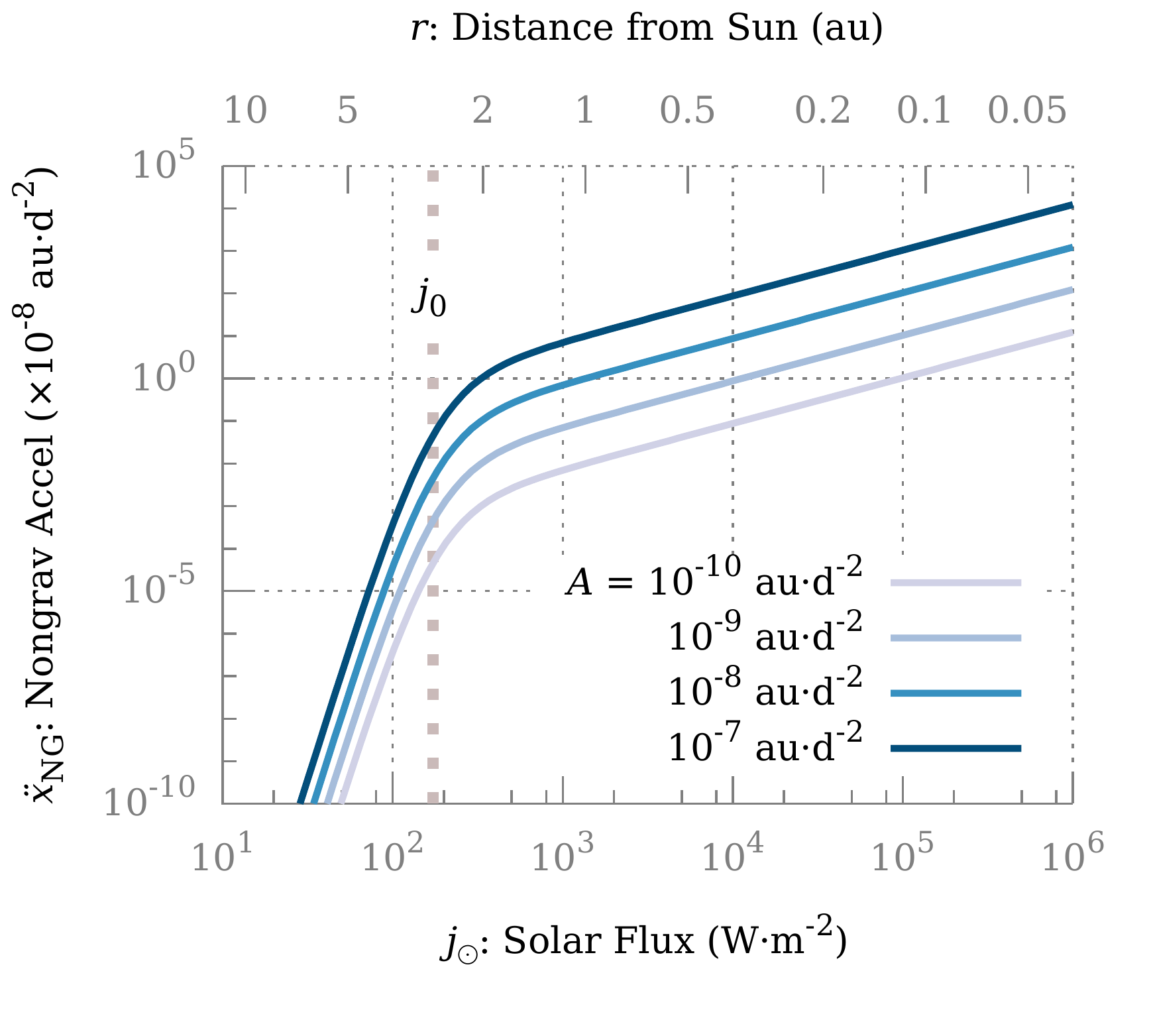}{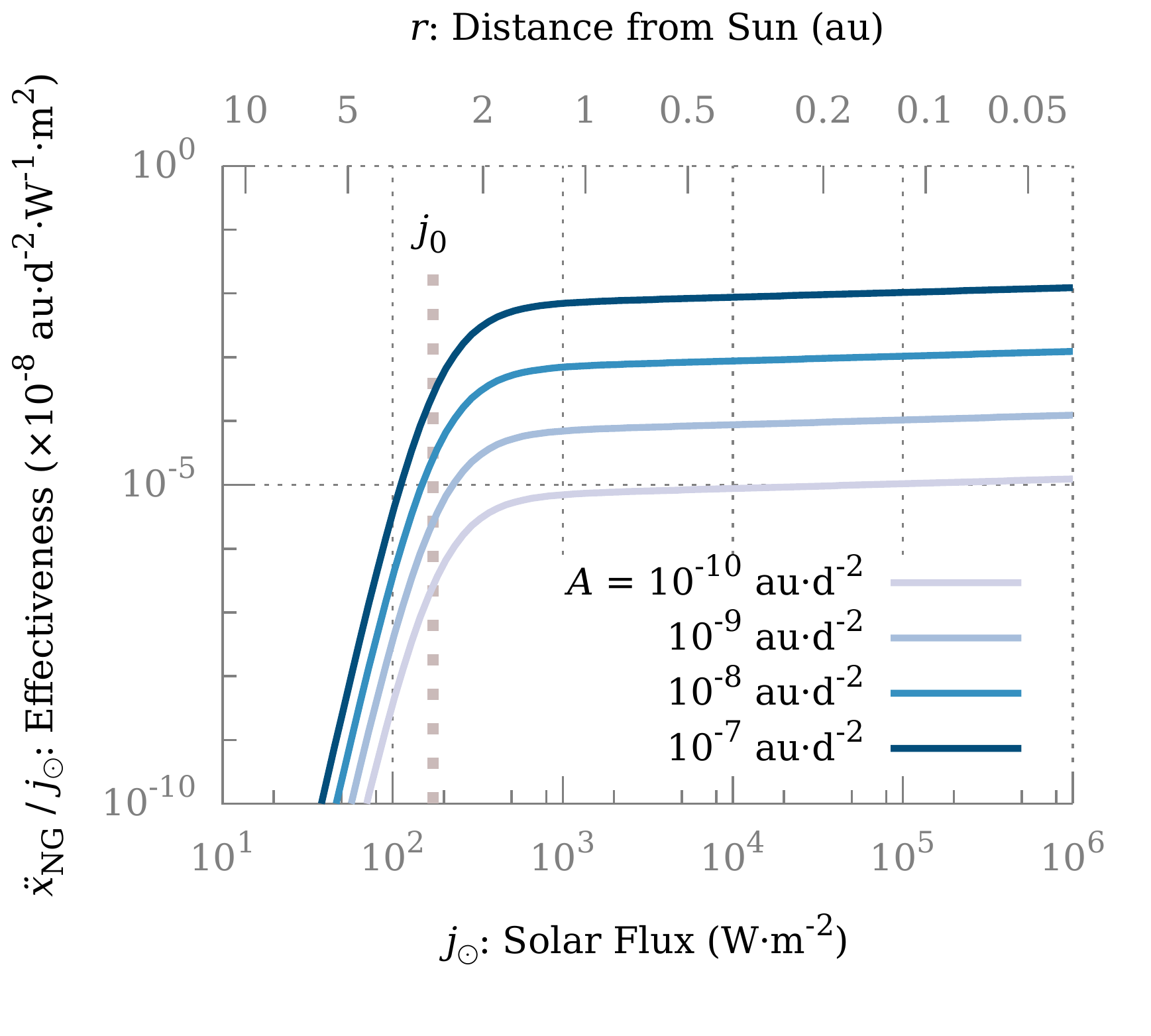}
\par\end{centering}
\caption{The nongravitational acceleration of a comet varies as a function
of incident flux (\textit{left}), and thus distance to the Sun, as
given by Eqs.~\ref{eq:sun-only} and \ref{eq:one-source}, based
on the model developed by \citet{marsden:1973}. A critical flux $j_{0}=172.6$~W~m$^{-2}$
divides the response into a highly nonlinear ($j\ll j_{0}$) regime
and a nearly linear ($j\gg j_{0}$) regime. Heating effectiveness
(\textit{right}), defined as the nongravitational acceleration per
unit of incident flux, approximately follows a step function centered
on $j_{0}$. \label{fig:com-nongrav}}
\end{figure*}

The simulations assume that energy absorption by the comet is wavelength-neutral,
i.e., that the nucleus is neutrally colored and any dust coma surrounding
the nucleus is optically thin. These conditions have thus far been
met for all comets visited by spacecraft to date, with the latter
condition likely met by all but a handful of comets with extremely
low perihelia \citep{gundlach2012}.

Under this assumption, the comet must necessarily respond equivalently
to all incident optical radiation, regardless of origin, with the
response depending only on the flux $j$ on the comet. By this ``equivalence
principle,''\textit{ }any\textit{ }radiation source at $\bm{x}_{0}$
(with $r'\equiv\|\bm{x}-\bm{x}_{0}\|$) uniformly illuminating the
cross section of the comet\textemdash including a laser with a beam
that has sufficiently diverged to a cross section larger than the
comet\textemdash will produce an acceleration

\begin{equation}
\ddot{\bm{x}}_{\text{NG}}=A\times\alpha\left(\frac{j}{j_{0}}\right)^{m/2}\left(1+\left(\frac{j}{j_{0}}\right)^{-n/2}\right)^{-k}\frac{\bm{x}-\bm{x}_{\odot}}{r'}\label{eq:one-source}
\end{equation}

where $j_{0}=172.6$~W~m$^{-2}$ is the solar flux at $r=r_{0}$,
given a solar irradiance of $S_{0}=1361$~W~m$^{-2}$ at $r=1$~au
\citep{kopp:2005}. The magnitude of single-source nongravitational
acceleration is plotted in Fig.~\ref{fig:com-nongrav} in the context
of the Sun. Two distinct regimes are evident:
\begin{enumerate}
\item Below a critical flux $j_{0}=172.6$~W~m$^{-2}$, acceleration falls
off rapidly as $\ddot{x}_{\text{NG}}\propto j^{12.83}$.
\item Above $j_{0}$, the function becomes nearly linear, approaching $\ddot{x}_{\text{NG}}\propto j^{1.075}$.
\end{enumerate}
The effectiveness of the heating\textemdash the amount of nongravitational
acceleration per unit of incident flux\textemdash is evidently closely
approximated by a step function separating the two regimes. Thus,
each unit of flux only contributes significantly to accelerating the
comet with total incident flux in the latter regime.

Eq.~\ref{eq:one-source}, however, only gives the acceleration from
a single radiation source. It is valid, for example, when the comet
is only being illuminated by the Sun, or is only being illuminated
by the laser. In a comet deflection scenario, both sources generally
must be considered. Because Eq.~\ref{eq:one-source} as a whole is
highly nonlinear, the acceleration from the superposition of the two
sources is nontrivial and requires additional assumptions regarding
the actual distribution of thrust over the comet's surface.

\begin{figure}
\begin{centering}
\plotone{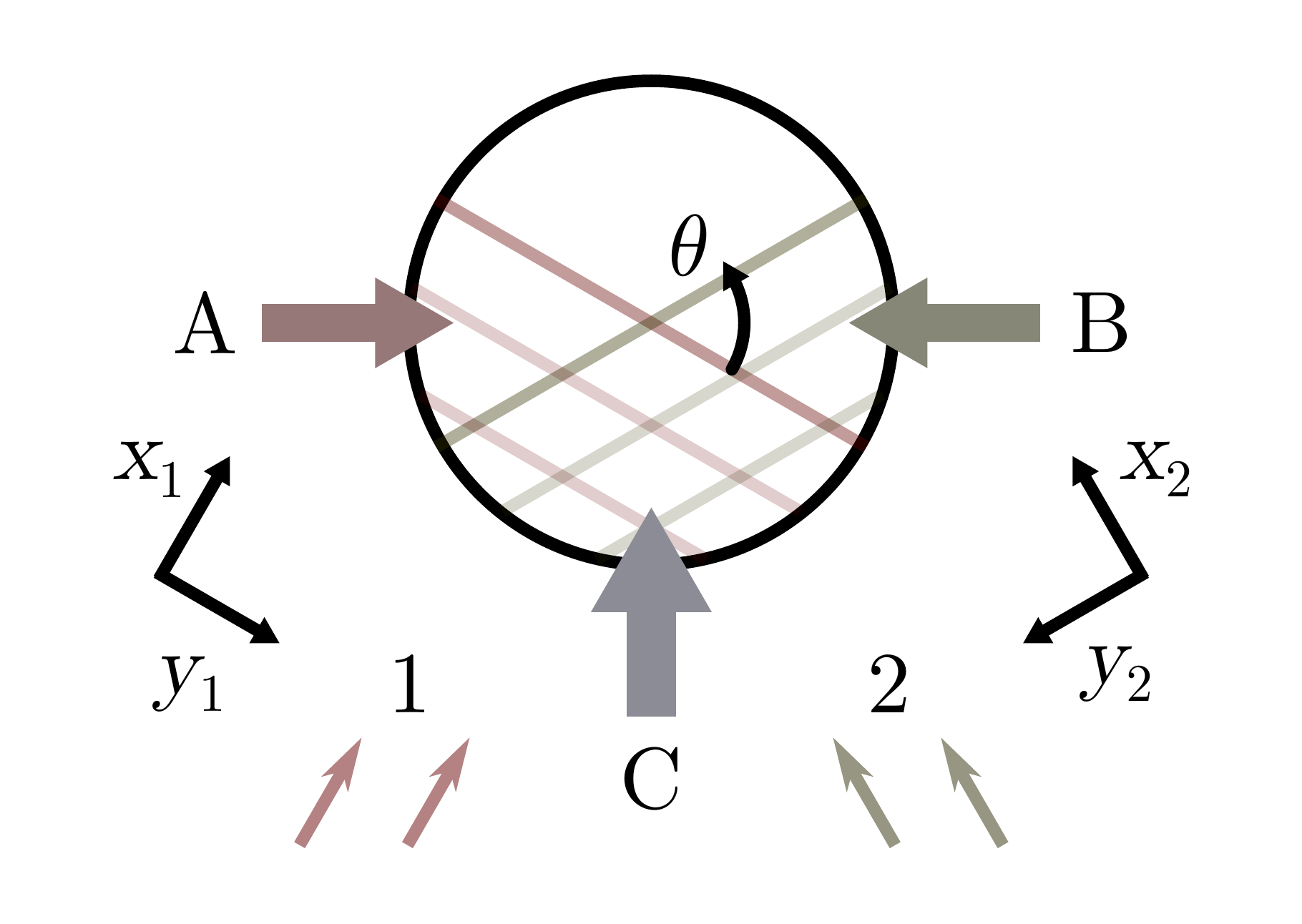}
\par\end{centering}
\caption{A comet being deflected is, in general, illuminated from two different
directions by two different radiation sources (the Sun and the laser).
In this diagram, source 1 illuminates the lower left (red stripes)
of the comet (circle) and source 2 illuminates the lower right (green
stripes), with the sources separated by an angle $\theta$. The illuminated
fraction of the comet is divided into three regions: region A is illuminated
by source 1 alone, region B is illuminated by source 2 alone, and
region C is illuminated by both. Arrows directed into the comet indicate
the assumed direction of acceleration contributed by each region\textemdash the
mean inward normal direction of the region\textemdash used in the
simulations. \label{fig:two-sources}}

\end{figure}

Consider two radiation sources 1 and 2, representing the Sun and the
laser, illuminating the comet and separated by angle $\theta$ as
illustrated in Fig. \ref{fig:two-sources}. The illuminated fraction
of the comet is divided into three regions:
\begin{enumerate}
\item Region A, illuminated by source 1 alone.
\item Region B, illuminated by source 2 alone.
\item Region C, illuminated by both.
\end{enumerate}
Due to the curvature of the comet's surface, the surface itself is
not uniformly illuminated in any of the three regions, despite the
cross section being uniformly illuminated. Precise determination of
the acceleration contributed by each region requires a detailed thermal
model for the surface response to incident radiation. Results from
such a model, which assumes a spherical comet, still only provide
a rough approximation for the acceleration of a realistic comet. Comparable
accuracy to a realistic comet may be attained by simply considering
the acceleration contributed by each region to be in the mean inward
normal direction of the region, as indicated in Fig.~\ref{fig:two-sources}.

We first select $\hat{\bm{x}}_{1}$ to be the propagation direction
of radiation from source 1, and $\hat{\bm{y}}_{1}$ a perpendicular
direction in the plane of both sources and the comet, as indicated
in Fig.~\ref{fig:two-sources}. When source 2 is inactive (i.e.,
no laser), the two-source model\textemdash the sum of the accelerations
contributed by region A and region C\textemdash must be consistent
with the single source model. Let $\ddot{\bm{x}}_{A}$ be the acceleration
contributed by region A and $\ddot{\bm{x}}_{C}$ be the acceleration
by region C. The sum $\ddot{\bm{x}}_{1}\equiv\ddot{\bm{x}}_{A}+\ddot{\bm{x}}_{C}$
must match the expression for $\ddot{\bm{x}}_{\text{NG}}$ given in
Eq.~\ref{eq:one-source}. Matching the components in $\hat{\bm{x}}_{1}$
and $\hat{\bm{y}}_{1}$ gives

\begin{equation}
\begin{cases}
\ddot{x}_{1} & =\ddot{x}_{A}\sin(\theta/2)+\ddot{x}_{C}\cos(\theta/2)\\
0 & =\ddot{x}_{A}\cos(\theta/2)-\ddot{x}_{C}\sin(\theta/2)
\end{cases}
\end{equation}

so $\ddot{x}_{A}=\ddot{x}_{1}\sin(\theta/2)$ and $\ddot{x}_{C}=\ddot{x}_{1}\cos(\theta/2)$
are the magnitudes of the acceleration contributions of the two regions.

When source 2 is activated, region A experiences no change, so $\ddot{x}_{A}$
remains unaffected. By symmetry, region B contributes an acceleration
of $\ddot{x}_{B}=\ddot{x}_{2}\sin(\theta/2)$, where $\ddot{\bm{x}}_{2}$
is the acceleration given by Eq.~\ref{eq:one-source} for source
2 alone. Finally, the acceleration contributed by region C becomes
roughly $\ddot{x}_{C}=\ddot{x}_{1+2}\cos(\theta/2)$, where $\ddot{\bm{x}}_{1+2}$
is the acceleration by Eq.~\ref{eq:one-source} for a single source
with the combined flux of both source 1 and source 2. The net nongravitational
acceleration on the comet is then the vector sum

\begin{equation}
\ddot{\bm{x}}_{\text{NG}}=\ddot{\bm{x}}_{A}+\ddot{\bm{x}}_{B}+\ddot{\bm{x}}_{C}\label{eq:two-sources}
\end{equation}

This two-source model degenerates into special cases of the one-source
model as expected in both the $\theta\to0$ limit (i.e., comet at
distance $r\gg1$~au, the separation of the Sun and the Earth/laser),
where $\ddot{\bm{x}}_{\text{NG}}\to\ddot{\bm{x}}_{1+2}$, and in the
$\theta\to\pi$ limit (i.e., comet directly between Sun and laser)
where $\ddot{\bm{x}}_{\text{NG}}\to\ddot{\bm{x}}_{1}+\ddot{\bm{x}}_{2}$,
a simple superposition of the one-source accelerations.

In the simulations, source 1 is the Sun, with an incident flux $j_{1}=S_{0}\times(\text{1 au}/r)^{2}$.
Source 2 is the laser, at distance $\Delta\equiv\|\bm{x}-\bm{x}_{\oplus}\|$
from the comet, producing a spot of radius $R_{\text{spot}}=\theta_{\text{beam}}\Delta$
with flux $j_{2}=P_{\text{peak}}/\left(\pi R_{\text{spot}}^{2}\right)$
when active.

The two-source model above is only valid when $R_{\text{spot}}>R_{\text{com}}$,
i.e., when the cross section of the comet is uniformly illuminated.
In the limit $j_{2}\gg j_{1}$ and $R_{\text{spot}}\ll R_{\text{com}}$
(but with $R_{\text{spot}}$ still sufficiently large to neglect thermal
diffusion across the surface\textemdash a condition assumed to always
hold), the laser-contributed acceleration decouples from the solar
acceleration $\ddot{\bm{x}}_{\odot}$ in Eq.~\ref{eq:sun-only} to
give

\begin{equation}
\ddot{\bm{x}}_{\text{s}\ll\text{c}}=\ddot{\bm{x}}_{\odot}+\left(R_{\text{spot}}/R_{\text{com}}\right)^{2}\ddot{\bm{x}}_{2}\label{eq:spot}
\end{equation}

where $\ddot{\bm{x}}_{2}$ is the one-source acceleration by a laser
of the same flux $j_{2}$ illuminating the entire comet cross section.

Note that the scaling relation in Eq.~\ref{eq:spot} assumes that
the rotation-averaged heating response of the comet is uniform at
the scale of the laser spot. Small-scale variations in terrain may
cause the net thrust to be directed in an unexpected direction, challenging
the earlier assumption of a dominant radial component of nongravitational
acceleration. This problem can be corrected by dithering the position
of the laser spot on the comet, which will average over these variations.

For intermediate $R_{\text{spot}}<R_{\text{com}}$ but $R_{\text{spot}}\not\ll R_{\text{com}}$,
linear interpolation (in area) between the $R_{\text{spot}}\to0$
limit and the case $R_{\text{spot}}\to R_{\text{com}}$ with $j_{2}\to j_{2}'=j_{2}\left(R_{\text{spot}}/R_{\text{com}}\right)^{2}$
is used. Total nongravitational acceleration by the Sun and laser
is therefore

\begin{equation}
\ddot{\bm{x}}_{\text{\ensuremath{\odot},las}}=\begin{cases}
\left(1-\left(R_{\text{spot}}/R_{\text{com}}\right)^{2}\right)\ddot{\bm{x}}_{\text{s}\ll\text{c}}\\
\:+\left(R_{\text{spot}}/R_{\text{com}}\right)^{2}\ddot{\bm{x}}_{\text{NG}}(j_{1},j_{2}'), & R_{\text{spot}}<R_{\text{com}}\\
\ddot{\bm{x}}_{\text{NG}}(j_{1},j_{2}), & R_{\text{spot}}\geq R_{\text{com}}
\end{cases}\label{eq:sun-laser}
\end{equation}

When $\langle P\rangle<P$, the laser is idle for a fraction of time
and $\ddot{\bm{x}}_{\text{NG}}$ becomes an appropriate linear combination
of $\ddot{\bm{x}}_{\text{\ensuremath{\odot\ }only}}$ from Eq.~\ref{eq:sun-only}
with the Sun alone, and $\ddot{\bm{x}}_{\text{\ensuremath{\odot},las}}$
from Eq.~\ref{eq:sun-laser} with the Sun and laser together:

\begin{equation}
\ddot{\bm{x}}_{\text{NG}}=\left(1-\langle P\rangle/P\right)\ddot{\bm{x}}_{\text{\ensuremath{\odot\ }only}}+\left(\langle P\rangle/P\right)\ddot{\bm{x}}_{\text{\ensuremath{\odot},las}}\label{eq:nongrav}
\end{equation}

Finally, there may be times when it is not advantageous to keep the
laser active, even when line of sight and power restrictions permit,
as perturbations to the orbit from laser heating at one part of the
orbit may cancel perturbations from laser heating at a different part
of the orbit \citep{zhang:2016:pasp}. Perturbation cancellation may
be minimized by tracking the sign of $\xi\equiv(\bm{x}_{\text{com}}-\bm{x}_{\oplus})\cdot\dot{\bm{x}}_{\text{com}}$
and permitting the laser to activate either only when $\xi>0$ (laser
is ``behind'') or only when $\xi<0$ (laser is ``ahead''). The
simulations focus primarily on deflecting comets with long orbital
periods $\gtrsim$100~yr, where deflection occurs only over the final
fraction of an orbit before its Earth encounter. Thus, $\xi<0$ nearly
always holds, and so the laser ``ahead'' condition is chosen.

\subsection{Numerical Setup}

The original orbit of the comet is specified by its perihelion distance
$q$, inclination $i$, eccentricity $e$, time of impact $T$, whether
impact occurs at the ascending or descending node, and whether impact
occurs before or after perihelion.

Next, initial conditions for the comet are found by the following
procedure:
\begin{enumerate}
\item Choose $\bm{x}_{0}(T)=\bm{x}_{\oplus}(T)$ as the final position of
the comet in its natural orbit.
\item Compute $\dot{\bm{x}}_{0}(T)$ such that the heliocentric Keplerian
orbit fit through $\bm{x}_{0}(T)$, $\dot{\bm{x}}_{0}(T)$ matches
the specified orbital parameters.
\item Using the Keplerian orbit of the comet, find the smallest $\delta t>0$
such that $\|\bm{x}_{0}(T-\delta t)-\bm{x}_{\oplus}(T-\delta t)\|=R_{\oplus}$,
the radius of the Earth.
\item Increase $\dot{x}_{0}(T-\delta t)\to\dot{x}_{0}(T-\delta t)+\sqrt{2GM_{\oplus}/R_{\oplus}}$
to account for Earth's gravitational well, where $M_{\oplus}$ is
the mass of the Earth.
\item Numerically integrate time-reversed system in the solar system potential
with $\ddot{\bm{x}}_{\text{NG}}$ from Eq.~\ref{eq:sun-only} to
find $\bm{x}(t_{0})$, $\bm{\dot{x}}(t_{0})$, the state vector at
time $t=t_{0}$ when the laser is to be first activated.
\end{enumerate}
Using $\bm{x}(t_{0})$, $\bm{\dot{x}}(t_{0})$ as initial conditions,
numerical integration proceeds using the same solar system potential,
with $\ddot{\bm{x}}_{\text{NG}}$ from Eq.~\ref{eq:nongrav}. The
system is integrated either to $t=T$ (yielding $\bm{x}(T)$, $\bm{\dot{x}}(T)$)
or until $\Delta(t)\equiv\|\bm{x}(t)-\bm{x}_{\oplus}(t)\|<R_{\oplus}$
where the comet impacts the Earth.

\section{Results}

For each comet deflection scenario, a deflection distance $\Delta_{\text{def}}$
quantifies the effectiveness of the deflection. For comets with a
local minimum $\Delta_{\text{min}}=\Delta(t_{\text{min}})>R_{\oplus}$
(no impact), use $\Delta_{\text{def}}=\Delta_{\text{min}}$. Otherwise,
define the true time of impact $t_{\text{imp}}$ as $\Delta(t_{\text{imp}})=R_{\oplus}$
and $\dot{\Delta}(t_{\text{imp}})<0$. Deflection distance $\Delta_{\text{def}}$
is then defined as the corresponding $\Delta_{\text{min}}$ for the
trajectory $\bm{x}_{1}(t)$ with $\bm{x}_{1}(t_{\text{imp}})=\bm{x}(t_{\text{imp}})$
and $\dot{\bm{x}}_{1}(t)=\dot{\bm{x}}(t_{\text{imp}})$, the linear
continuation of the comet's trajectory through the Earth.

A typical comet discovered in the near future might follow a timeline
similar to that of the recent dynamically new comet C/2013 A1 (Siding
Spring), which passed Mars at a distance of 140500~km (0.37~LD)
in 2014 October, just 22 months after discovery \citep{farnocchia:2016}.
Continuing advancements in survey capability may conceivably extend
the advance notice by a few months to years in the near future, though
detection of such comets is ultimately limited by their trajectories,
approaching from the distant outer solar system.

A future Earth-bound comet might be discovered $\sim$3~yr in advance,
permitting impact confirmation and laser activation by 2~yr prior
to the Earth encounter. For the simulations, consider a similar $\unit[2R_{\text{com}}=]{500}$~m
diameter comet with $A=2\times10^{-8}$~au~d$^{-2}$ ($A_{\unit[1]{km}}=1\times10^{-8}$~au~d$^{-2}$)
in a comparable orbit of $q=0.9$~au, $e=1$, $i=130\degree$ leading
to an Earth impact at $T=\text{J2000.0}$ at its ascending node while
the comet is inbound. These parameters for this canonical\textit{
}comet are used for all simulations, except when otherwise noted.

Note that the assumed value of $A_{\unit[1]{km}}=1\times10^{-8}$~au~d$^{-2}$
is lower than that of typical LPCs with $A_{\unit[1]{km}}\sim10^{-7}$~au~d$^{-2}$
or even $A_{\unit[1]{km}}\sim10^{-6}$~au~d$^{-2}$ \citep{krolikowska:2004}.
Simulation results will therefore underestimate deflection distance
for these more responsive comets by a corresponding factor of 1\textendash 2
orders of magnitude. Meanwhile, periodic comets (JFCs and HTCs) vary
in their composition, due to variation in dynamical age, and may have
comparable $A_{\unit[1]{km}}\sim10^{-8}$~au~d$^{-2}$ or lower,
with nongravitational deflection often not detectable at all if volatiles
are sufficiently depleted. The presented results focus on the case
of LPC impactors, which are associated with the extremely short warning
times and cannot be reliably be deflected by any other proposed method.
For these cases, the canonical comet described above provides an adequate,
conservative example.

\subsection{Orbital Laser}

A laser array in Earth orbit is restricted in $\langle P\rangle$
by the size and efficiency of its PV array. Consider a square PV array
with edge length $L_{\text{las}}$, equal in size to the laser array.
For a total solar-to-laser power efficiency $\varepsilon=20\%$, such
a system produces $\langle P\rangle=\varepsilon S_{0}L_{\text{las}}^{2}$.
With $\varepsilon$ constrained by technology and thermodynamics,
$\langle P\rangle$ can only reliably be improved by scaling up the
array. Use of a supplementary battery system, however, allows $P\gg\langle P\rangle$.

\begin{figure*}
\begin{centering}
\plottwo{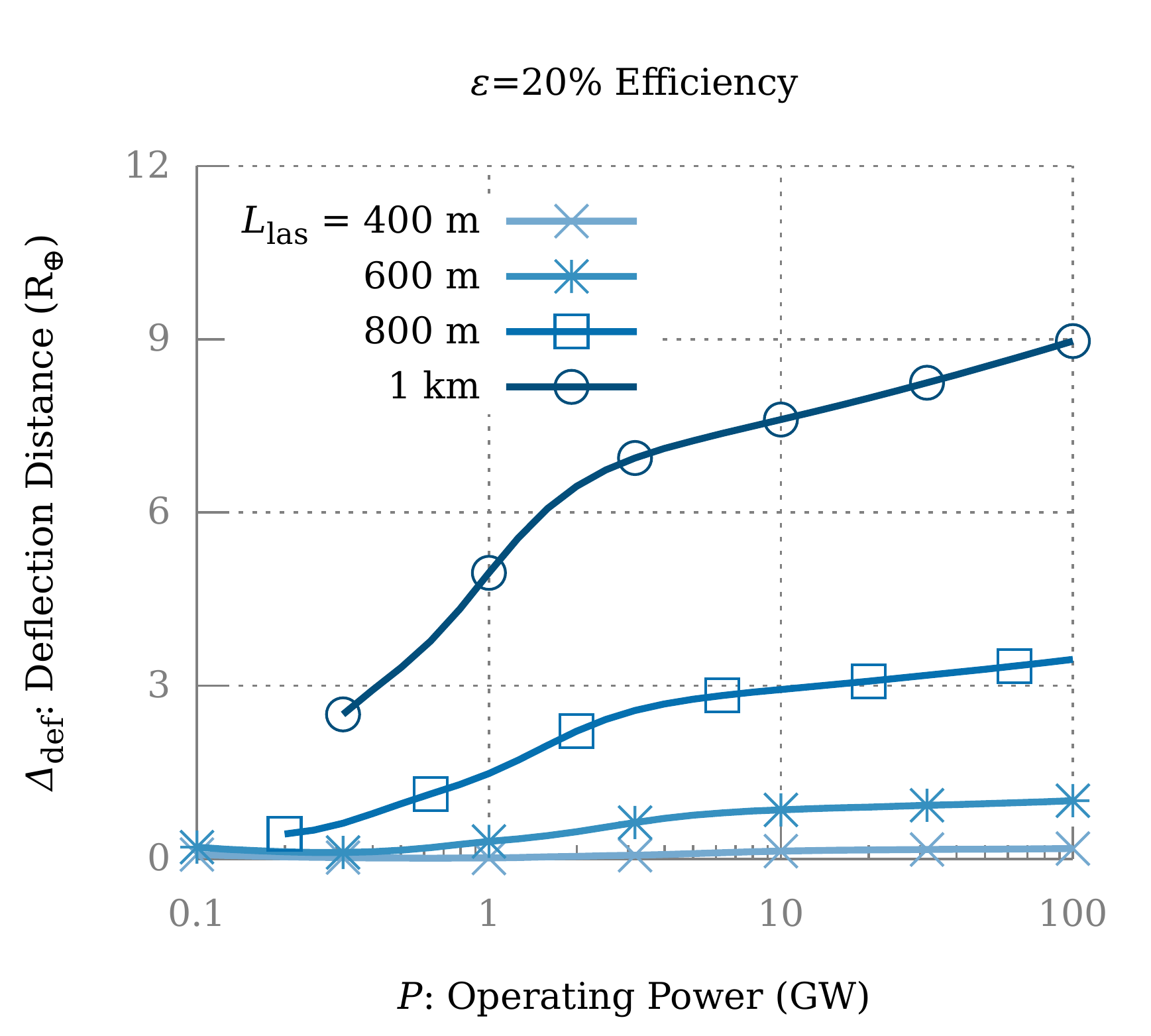}{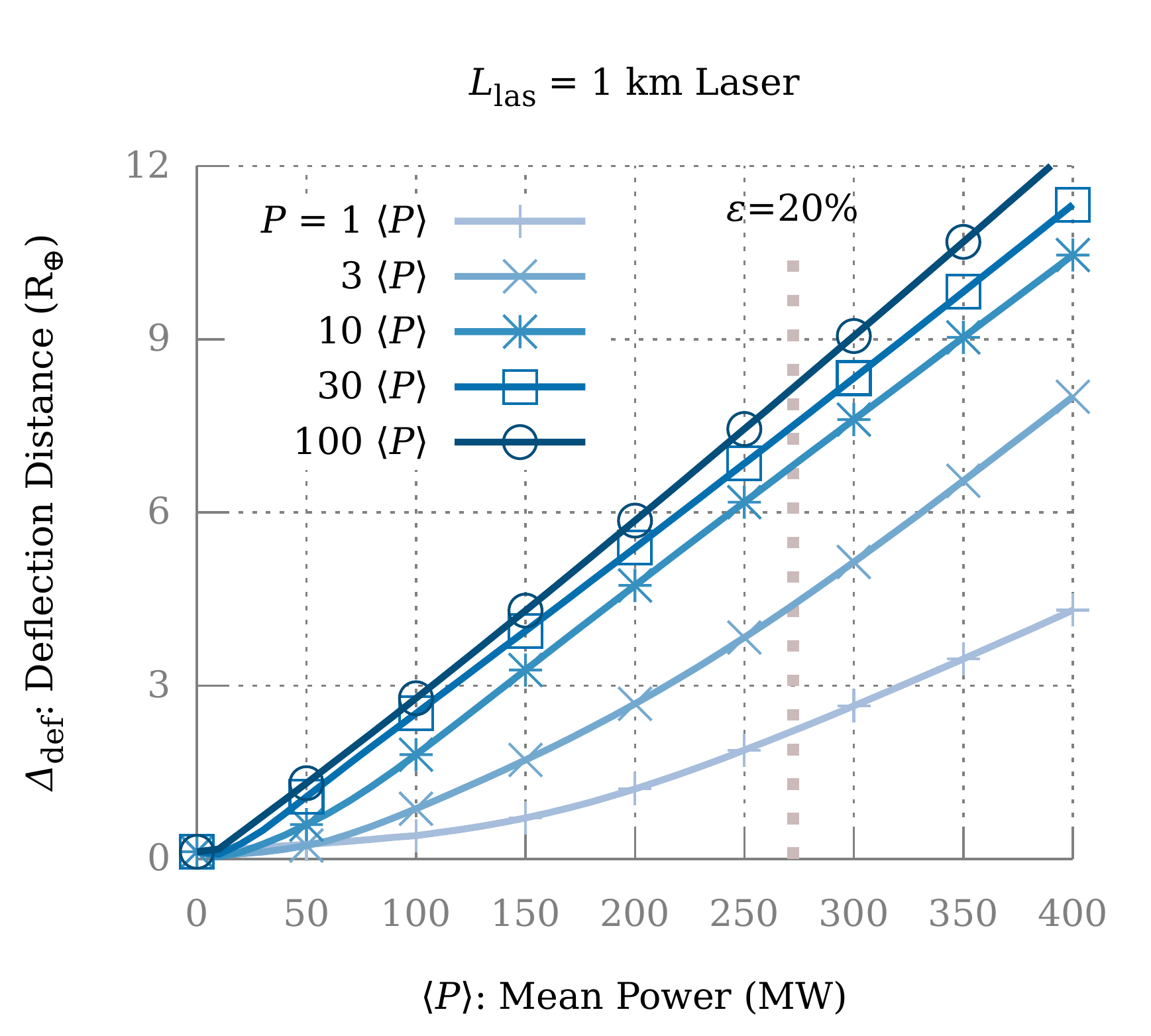}
\par\end{centering}
\caption{Deflection of the canonical comet by orbital laser arrays over 2~yr:
Increasing $P$, even with a fixed $\langle P\rangle$, yields a substantial
improvement in deflection due to the nonlinear heating response of
the comet illustrated in Fig. \ref{fig:com-nongrav}. Note that $P\protect\geq\langle P\rangle$,
so curves are terminated at the lower end $P=\langle P\rangle$ (\textit{left})
by this condition. \label{fig:orbit-power}}
\end{figure*}

Fig.~\ref{fig:orbit-power} illustrates the effectiveness of arrays
with a range of $L_{\text{las}}$, $\langle P\rangle$ and $P$ for
deflecting the canonical comet over 2~yr. Increasing array size $L_{\text{las}}$
and efficiency $\varepsilon$ both yield a substantial improvement
in deflection distance $\Delta_{\text{def}}$. Furthermore, an increase
in $P$ alone leads to a significant increase in $\Delta_{\text{def}}$.
This result illustrates the nonlinear behavior of Eq.~\ref{eq:sun-only},
which makes each unit of incident flux much more effective at accelerating
the comet at $j\gg j_{0}$ than at $j\ll j_{0}$, as shown in Fig.~\ref{fig:com-nongrav}.
Increasing $P$ extends the range over which the comet can be illuminated
at $j\gg j_{0}$, enabling a greater deflection distance for the same
amount of energy. Note, however, that when $P$ is sufficiently high,
the comet will be illuminated at $j\gg j_{0}$ for the entirety of
the deflection, and further increases in $P$ will have little effect
on heating effectiveness, and thus, $\Delta_{\text{def}}$.

It is conceivable that comet detection capability advances sufficiently
by the time a threatening comet is identified that deflection may
begin as early as $t_{0}=T-5$~yr for larger comets, which may permit
the use of a smaller laser array. However, at such an early time,
the comet is a large distance $r\text{, }\Delta\sim15$~au from both
the Sun and the laser. Without a sufficiently high operating power,
the flux on the comet will fall deep within the $j\ll j_{0}$ regime
and little deflection will occur until the comet approaches to a much
closer distance.

\begin{figure}
\begin{centering}
\plotone{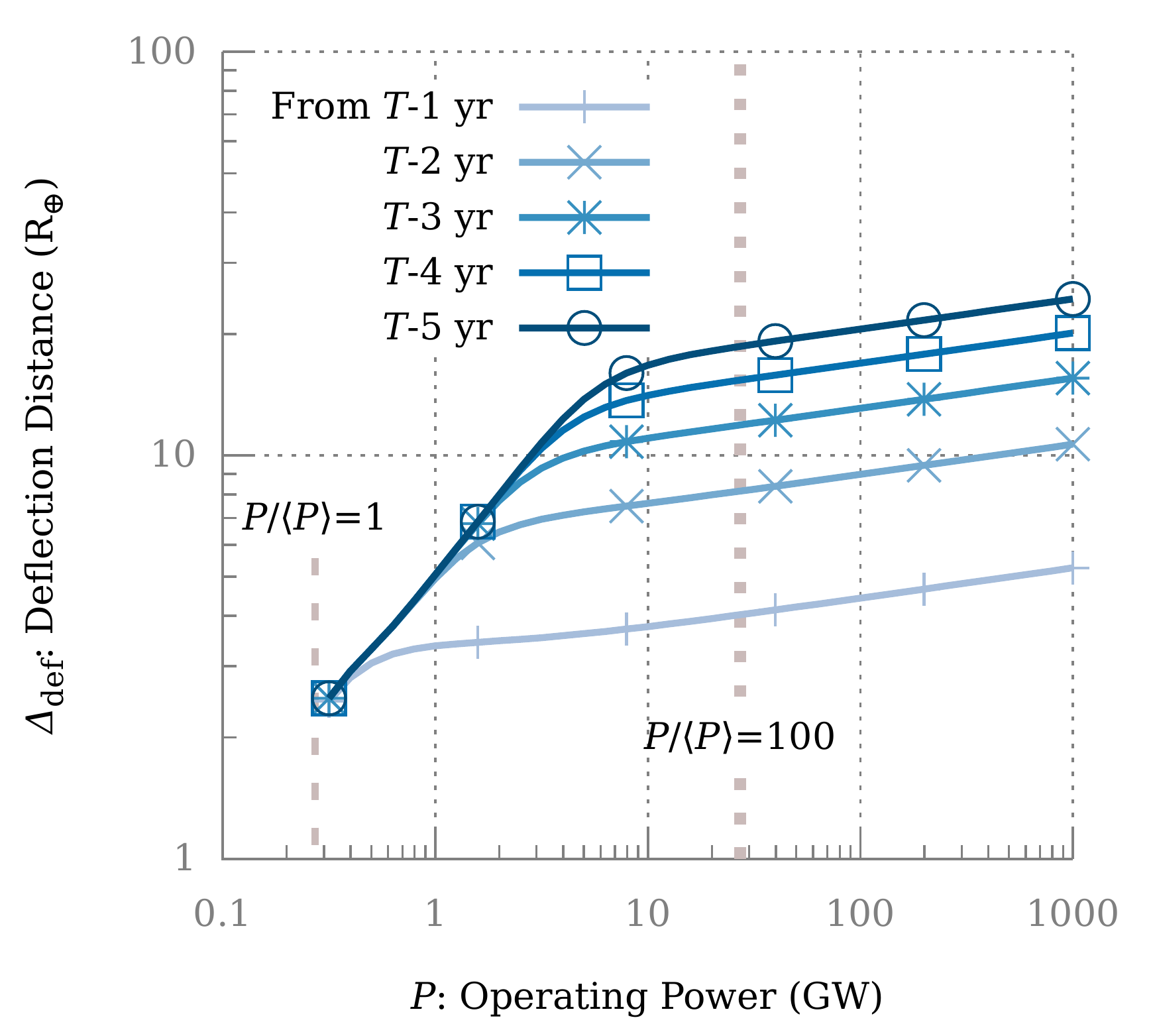}
\par\end{centering}
\caption{Effectiveness of a $L_{\text{las}}=1$~km laser with $\varepsilon=20\%$
operating at various $P$: Starting deflection earlier is only of
measurable benefit if $P$ is sufficient high for $j\gg j_{0}$ on
the comet to give non-negligible acceleration at the beginning of
the deflection period. \label{fig:orbit-early}}
\end{figure}

Fig.~\ref{fig:orbit-early} compares the effectiveness of several
deflection start times for a smaller $L_{\text{las}}=500$~m laser
with $\varepsilon=20\%$ efficiency. Operating at $P=100\langle P\rangle=7$~GW,
the canonical comet is deflected $\Delta_{\text{def}}\sim20$~$R_{\oplus}$
when deflection begins at $T-2$~yr. Beginning deflection even earlier
achieves little additional gain in deflection, in the absence of an
additional increase in $P$.

\begin{figure}
\begin{centering}
\includegraphics[width=1\linewidth]{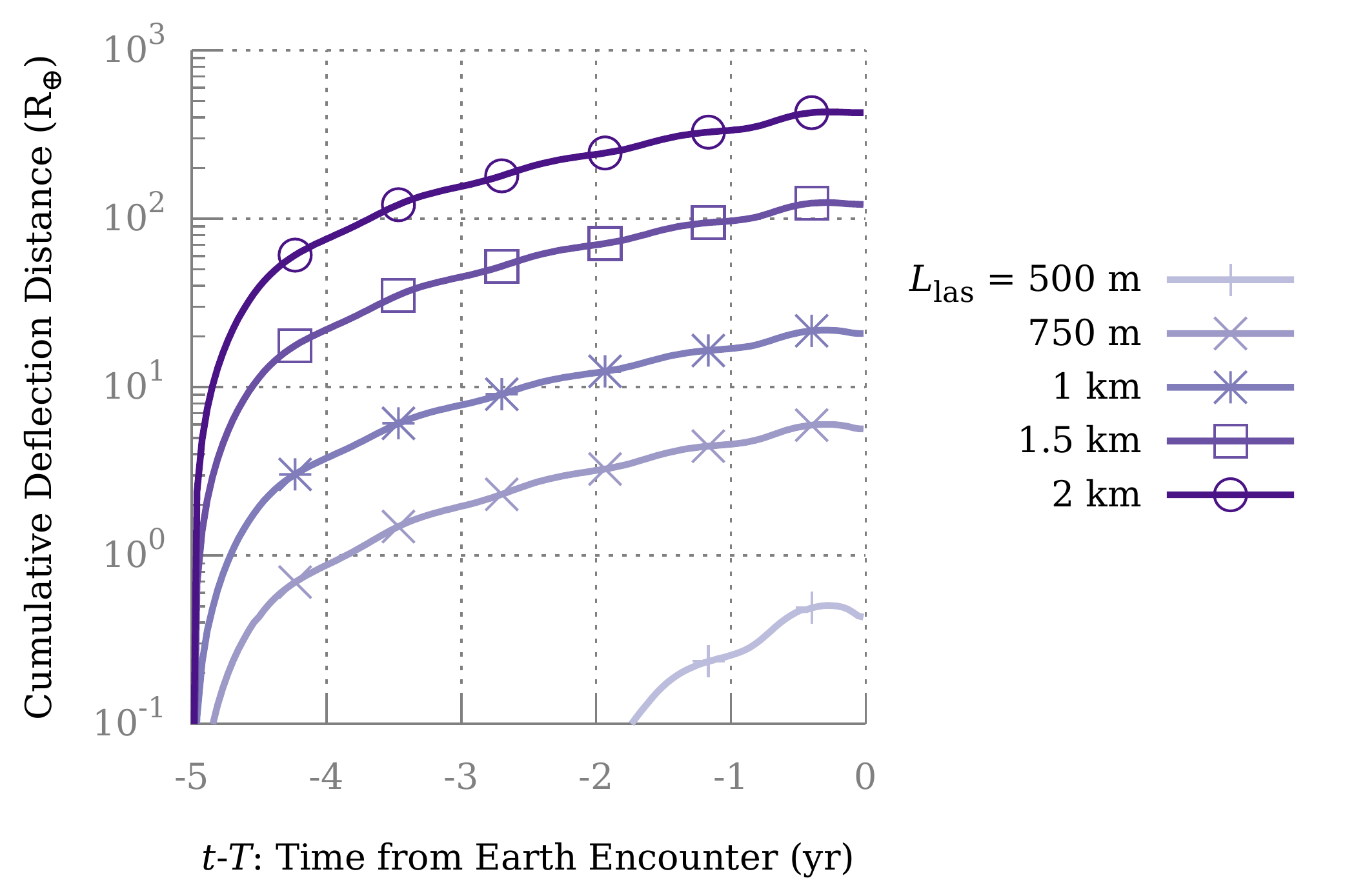}
\par\end{centering}
\caption{Deflection of the canonical $2R_{\text{com}}=500$~m comet is illustrated
for laser arrays of various sizes operating at $P/\langle P\rangle=100$
and $\varepsilon=20\%$ beginning at $t_{0}=T-5$~yr. This plot shows
the cumulative deflection distance at time $t$\textemdash the deflection
distance $\Delta_{\text{def}}$ if the deflection process were to
terminate at time $t$\textemdash over the interval of deflection.
Note that deflection in the final months opposes the deflection accumulated
earlier, so maximal deflection is actually attained a few months before
the comet arrives at Earth. \label{fig:orbit-branches}}
\end{figure}

This effect is further illustrated by Fig.~\ref{fig:orbit-branches},
which shows the accumulation of deflection of a set of canonical comets
by laser arrays of various sizes operating at $P/\langle P\rangle=100$
and $\varepsilon=20\%$, beginning at $t=T-5$~yr. Over the first
3~yr, a $L_{\text{las}}=400$~m laser is incapable of deflecting
the comet by even 0.1~$R_{\oplus}$, as $j\ll j_{0}$ over this period.
The bulk of the eventual $\Delta_{\text{def}}=3.5$~$R_{\oplus}$
is accumulated over the final 1~yr. In contrast, the higher flux
attained by laser arrays of $L_{\text{las}}\gtrsim600$~m begins
to accumulate deflection immediately at $t=t_{0}=T-5$~yr. Note that
the final months of deflection contribute negatively to the final
$\Delta_{\text{def}}$. Optimal deflection requires terminating the
deflection process a few months before the comet arrives at Earth.
As the loss in the final $\Delta_{\text{def}}$ is generally under
1~$R_{\oplus}$, no attempt is made to precisely determine the optimal
cutoff time for these results, as this effect is dwarfed by the uncertainties
associated with comet deflection discussed earlier.

Note that a laser at $P/\langle P\rangle=100$ would operate for an
average of only 14.4~min each day, during which time the energy collected
over an entire day is drained. Achieving such high $P/\langle P\rangle$,
which requires a high-density laser array, while maintaining $\varepsilon$
may not necessarily be less of an engineering challenge than constructing
a larger and equally effective array at lower $P/\langle P\rangle$.
Analysis of optimal $P/\langle P\rangle$ is left to a more thorough
investigation of orbital laser array construction. The remainder of
this section considers arrays operating at $P/\langle P\rangle=100$.

\begin{figure*}
\begin{centering}
\plottwo{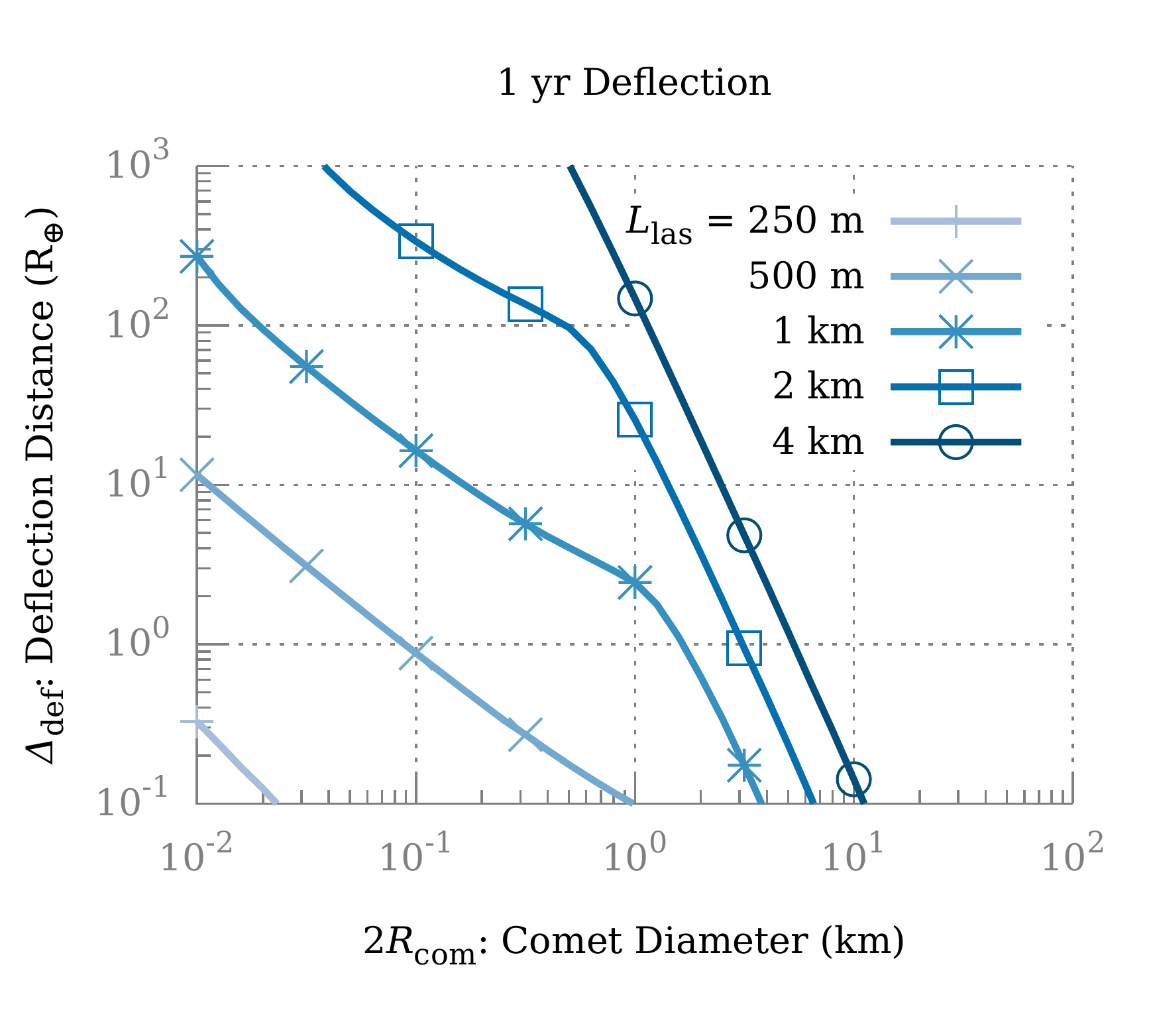}{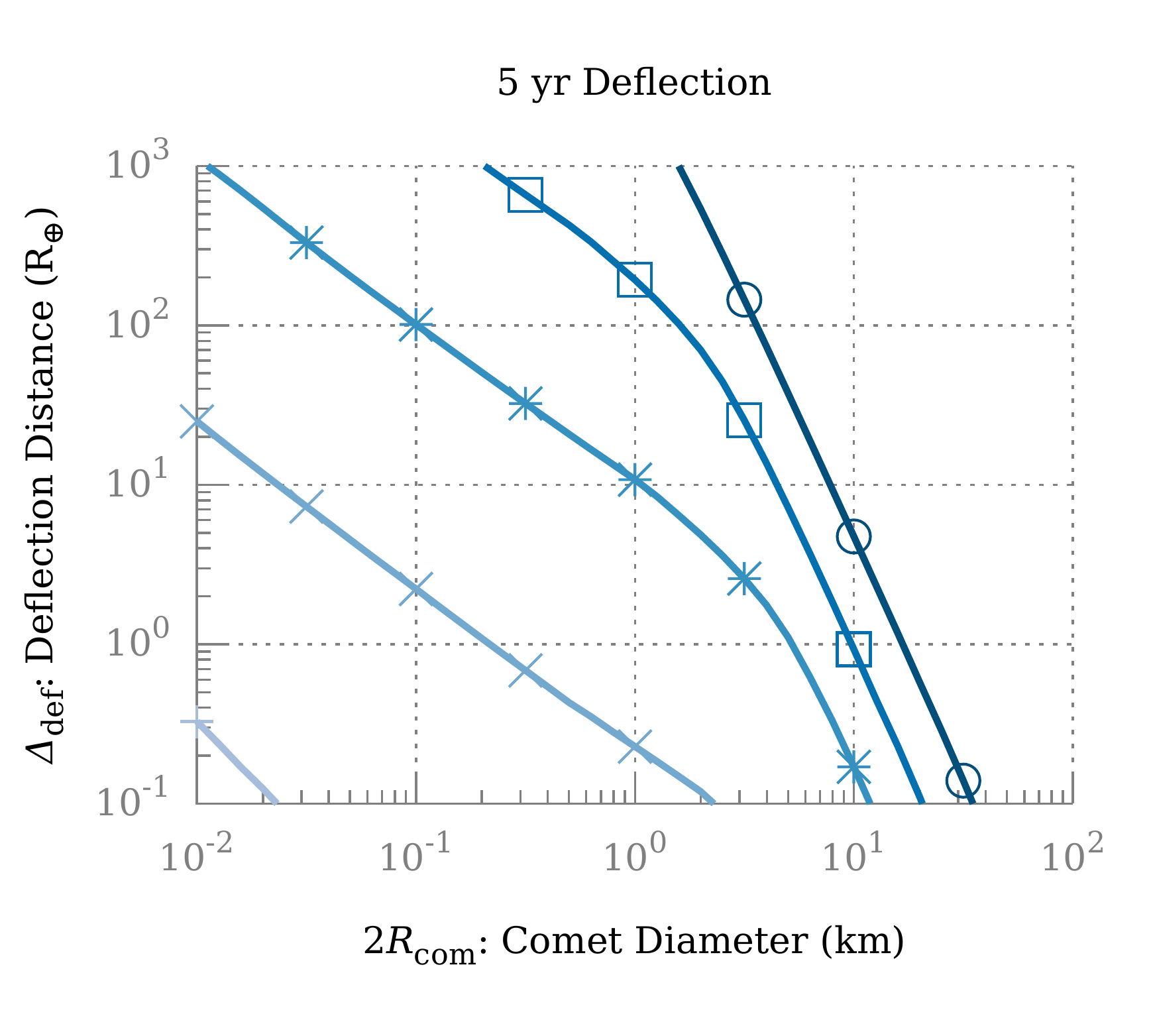}
\par\end{centering}
\caption{Under a fixed $A_{\unit[1]{km}}=1\times10^{-8}$~au~d$^{-2}$, deflection
distance $\Delta_{\text{def}}$ scales roughly as $\Delta_{\text{def}}\propto R_{\text{com}}^{-1}$
at small $R_{\text{com}}$, for which $R_{\text{spot}}>R_{\text{com}}$
during most of the deflection period, and as $\Delta_{\text{def}}\propto R_{\text{com}}^{-3}$
at large $R_{\text{com}}$, for which $R_{\text{spot}}<R_{\text{com}}$
during most of the deflection period. In addition, larger laser arrays,
which are needed to deflect larger comets, benefit more from longer
deflection periods. With 1~yr available for deflection (\textit{left}),
a $L_{\text{las}}=1$~km laser can deflect a $2R_{\text{com}}=1$~km
comet by $\Delta_{\text{def}}\approx2$~$R_{\oplus}$. Given 5~yr
(\textit{right}), the same laser can deflect a larger 3~km comet
by the same distance. Note that $\Delta_{\text{def}}\propto A_{\unit[1]{km}}$,
so for highly active LPCs with $A_{\unit[1]{km}}\sim10^{-7}$\textendash $10^{-6}$~au~d$^{-2}$,
these curves would shift upward by a corresponding 10\textendash $100\times$.
\label{fig:orbit-comdiam}}

\end{figure*}

A larger array or additional warning time is necessary to divert comets
of $2R_{\text{com}}>1$~km. Fig.~\ref{fig:orbit-comdiam} shows
that $\Delta_{\text{def}}\propto R_{\text{com}}^{-3}$ for a given
$A_{\text{\ensuremath{\unit[1]{km}}}}$. With 1~yr available for
deflection, a $L_{\text{las}}=1$~km laser can deflect a $2R_{\text{com}}=1$~km
comet by $\Delta_{\text{def}}=3$~$R_{\oplus}$. Given 5~yr, the
same laser can deflect a much larger $2R_{\text{com}}=3$~km comet
by the same distance. Very large comets of $2R_{\text{com}}>10$~km
require very large laser arrays of $L_{\text{las}}\gtrsim4$~km with
$\gtrsim$5~yr to safely deflect. It is important to remember that
all of the simulations assume the canonical $A_{\unit[1]{km}}=1\times10^{-8}$~au~d$^{-2}$.
Because $\Delta_{\text{def}}\propto A_{\unit[1]{km}}$, if these comets
behave similarly to volatile-rich LPCs with $A_{\unit[1]{km}}\sim10^{-7}$\textendash $10^{-6}$~au~d$^{-2}$,
deflection becomes a corresponding 10\textendash $100\times$ more
effective, and a $L_{\text{las}}=1$~km laser may be enough to deflect
such a 5\textendash 10~km comet in 1~yr or a 10\textendash 20~km
comet in 5~yr.

Deflection effectiveness drops rapidly with decreasing array size.
At $P/\langle P\rangle=100$, $L_{\text{las}}=500$~m appears to
be the smallest useful array for comet deflection, and is capable
of deflecting a canonical $2R_{\text{com}}=50$~m comet by $\Delta_{\text{def}}=10$~$R_{\oplus}$.
Note that, because the simulations assume $A$ and $R_{\text{com}}$
remain static throughout the deflection, results for small comets,
which are more strongly altered by the deflection process, should
be treated with caution.

\begin{figure*}
\begin{centering}
\plottwo{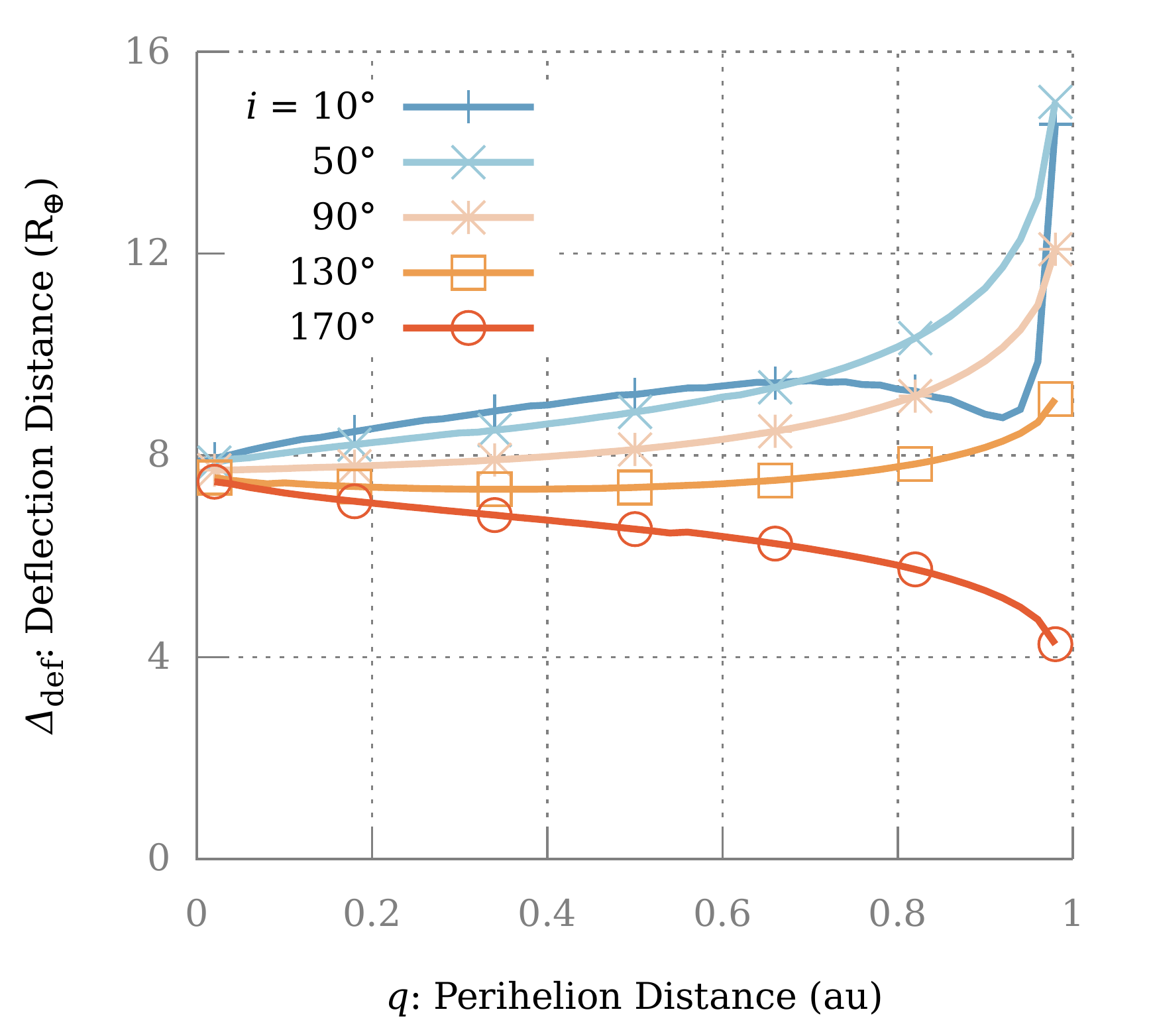}{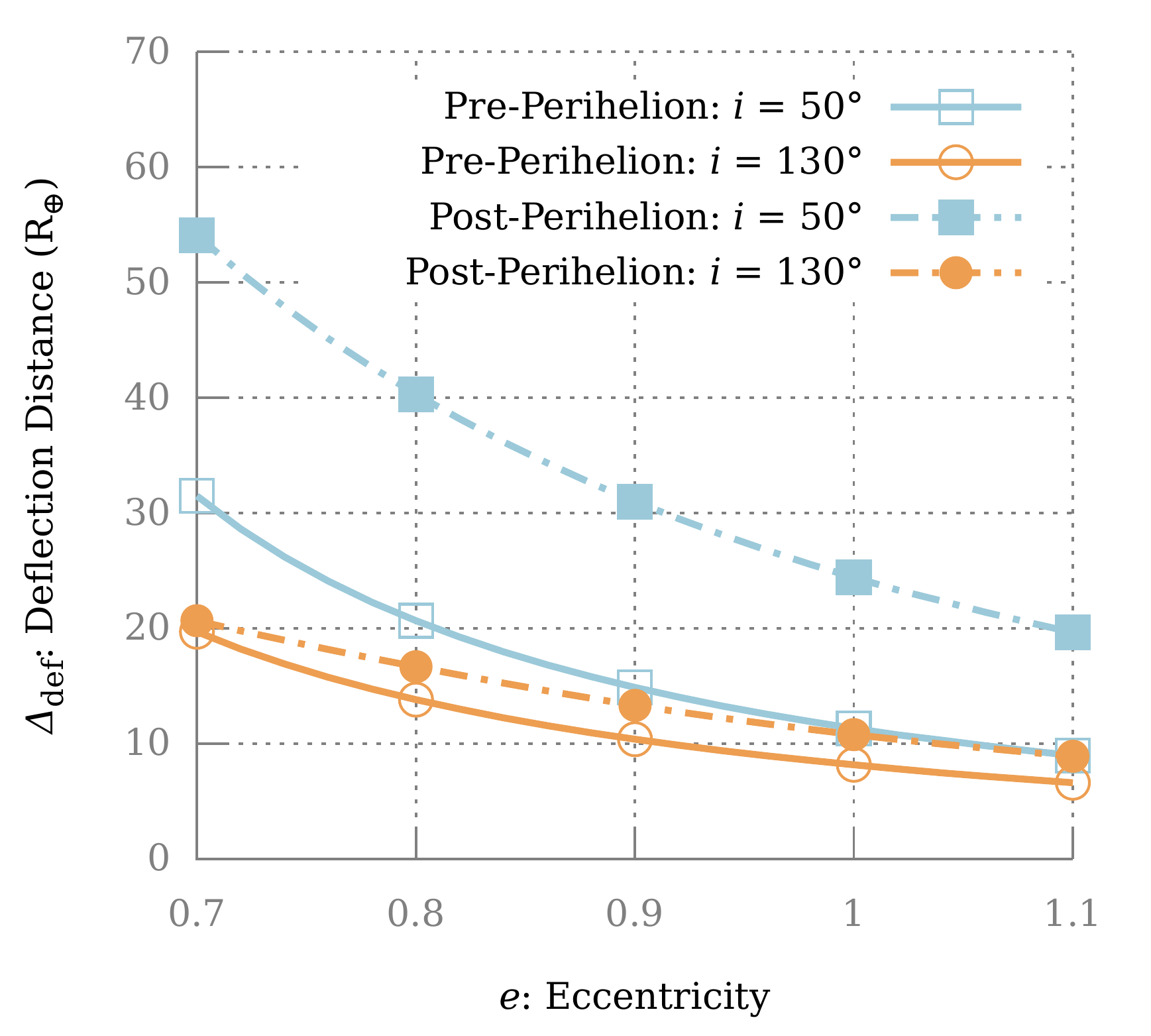}
\par\end{centering}
\caption{Deflection effectiveness of a $L_{\text{las}}=1$~km, $P=100\langle P\rangle$
orbital laser targeting copies of the canonical comet with modified
orbital elements over 2~yr. High-inclination prograde orbits are
most favorable to deflection, while low-inclination retrograde orbits
are least favorable \textit{(left}). Furthermore, a pre-perihelion
impact appears to be more difficult to mitigate than a post-perihelion
impact, particularly for comets in prograde orbits (\textit{right}).
\label{fig:orbit-vary}}

\end{figure*}

Finally, because the solar system is not gravitationally isotropic
about the Earth, the deflection effectiveness of a given laser system
varies depending on the exact orbit of the comet. Generally, deflection
is most effective for orbits that place the comet near the laser for
long durations over the deflection period, because $\ddot{x}_{\text{NG}}$
is an increasing function of $j\propto\Delta^{-2}$.

Fig.~\ref{fig:orbit-vary} shows the variations in effectiveness
for a $L_{\text{las}}=1$~km, $P=100\langle P\rangle$ laser deflecting
an otherwise canonical comet, beginning at $T-1$~yr over a range
of plausible orbital parameters. For this case, high-inclination prograde
orbits are most favorable and low-inclination retrograde orbits are
least favorable, from a deflection standpoint. Furthermore, an impact
while the comet is inbound is more difficult to mitigate than if the
impact occurred were while the comet is outbound. The latter phenomenon
is explained by the comet's final approach to Earth: a comet approaching
from within the Earth's orbit (post-perihelion encounter) approaches
more rapidly and spends less time near the Earth than an otherwise
identical comet approaching from beyond Earth's orbit (pre-perihelion
encounter). In all cases, the variation in effectiveness from orbital
differences is within a factor of 2\textemdash no more than the variation
in $A_{\unit[1]{km}}$ between dynamically similar comets \citep{yeomans:2004}.

Lasers with larger $L_{\text{las}}$ starting at earlier $t_{0}$
experience increasingly less variation between comets of different
orbits as deflection occurs over a spatial scale much larger than
Earth's orbit with $j>j_{0}$ over a much longer distance. At such
scale, the gravitational potential of the solar system is nearly isotropic
about the laser (which, at large scale, is located near the center
of the solar system) and deflection approaches the orbit-neutral limit.
Conversely, small lasers are affected more strongly by the orbit of
the comet, an effect that becomes important for ground-based lasers
which may be useful for deflection at much smaller scales.

\subsection{Terrestrial Laser}

Unlike the case for orbital arrays, $\langle P\rangle$ is not restricted
by $L_{\text{las}}$ for terrestrial laser arrays where electric power
may be supplied externally. For a given $P$, $\langle P\rangle$
is only restricted by the requirement that the comet be within the
laser's field of view $\Theta_{\text{fov}}$ and that weather conditions
permit operation. Achieving the necessary diffraction-limited beam
from the ground poses a serious challenge for very large $L_{\text{las}}$
with their tiny $\theta_{\text{beam}}$. These constraints favor compact
but high-powered arrays.

Terrestrial lasers are directionally biased by the nature of their
fixed field of view relative to Earth. A laser at latitude $\phi_{\text{las}}$
with field of view $\Theta_{\text{fov}}$ may only target comets in
declinations $\phi_{\text{las}}-\Theta_{\text{fov}}/2<\delta_{\text{com}}<\phi_{\text{las}}+\Theta_{\text{fov}}/2$.
A laser at a far northern latitude is completely ineffective against
a comet approaching from near the southern celestial pole, as such
a comet never rises sufficiently high in the sky to enter the laser's
field of view. 

\begin{figure*}
\begin{centering}
\plottwo{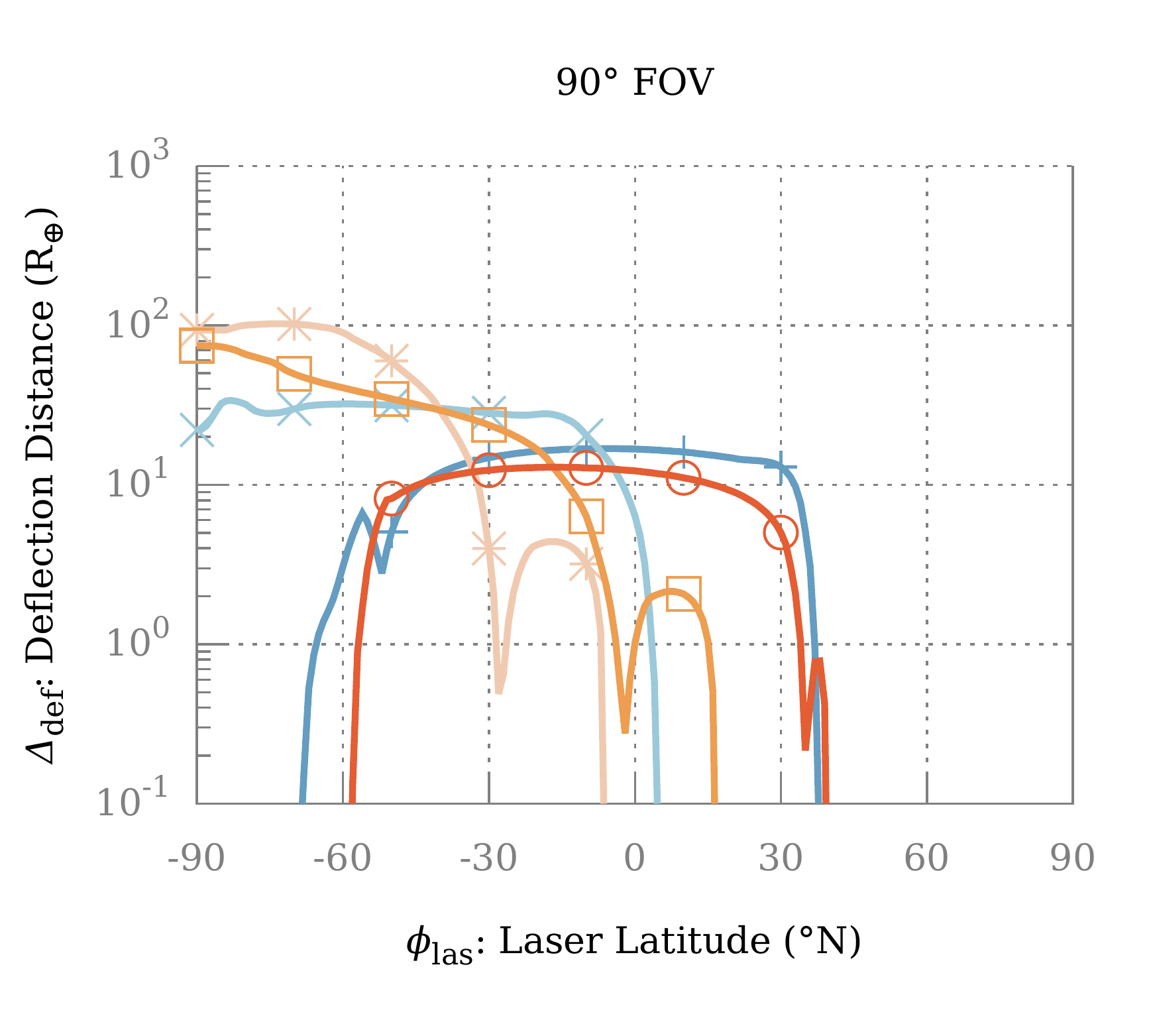}{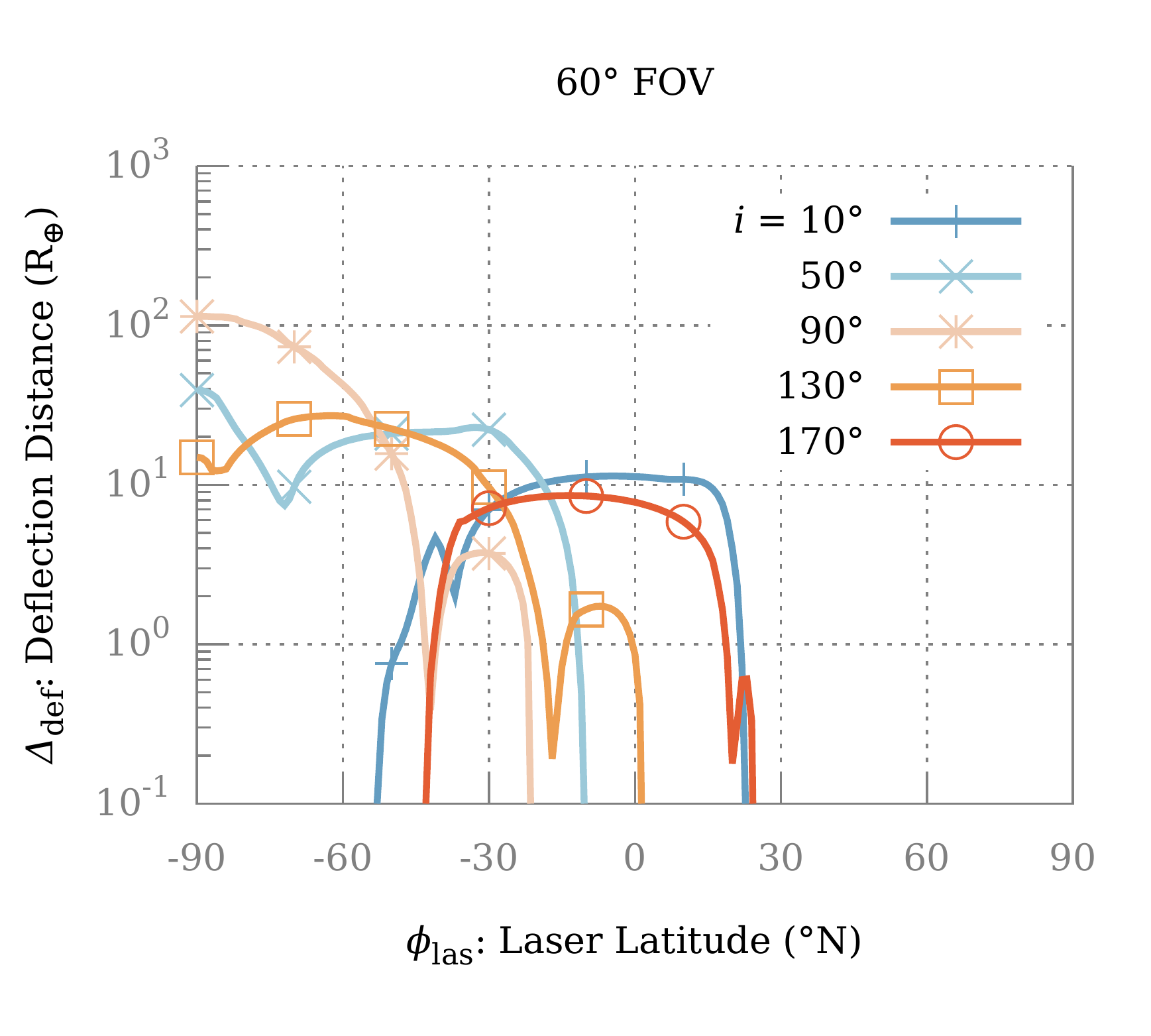}
\par\end{centering}
\caption{Deflection effectiveness of $L_{\text{las}}=500$~m, $P=25$~GW
terrestrial lasers located at various latitudes $\phi_{\text{las}}$
with identical $\kappa=50\%$ targeting copies of the canonical comet
of various inclinations $i$ over 2~yr: A laser with a $90\degree$
field of view (\textit{left}) is more effective against a wider range
of comets, than an otherwise identical laser with a $60\degree$ field
of view (\textit{right}). \label{fig:ground-lat}}

\end{figure*}

Fig.~\ref{fig:ground-lat} compares the deflection effectiveness
against a set of modified canonical comets of various $i$ by a $L_{\text{las}}=500$~m,
$P=25$~GW array at $\kappa=50\%$ for $\Theta_{\text{fov}}=90\degree$
and $\Theta_{\text{fov}}=60\degree$. The laser with the larger $90\degree$
field of view targets the comet longer than a laser with the smaller
$60\degree$ field of view, and thus it produces a larger deflection
$\Delta_{\text{def}}$ and is effective over a wider range of latitudes
$\phi_{\text{las}}$.

Prograde orbits ($i<90\degree$) are strongly favored over retrograde
($i>90\degree$), due to prograde-orbiting comets having slower relative
velocity in the final approach. The variations for the $L_{\text{las}}=500$~m
terrestrial laser are far more significant than those of the $L_{\text{las}}=1$~km
orbital laser, due to the spatial scale differences discussed in the
previous section. Note that these results are for an Earth encounter
at the comet's ascending node where the comet approaches from below
the ecliptic, favoring deflection from the southern hemisphere. Descending
node encounters correspond to similar results, but mirrored to favor
deflection from the northern hemisphere. \citet{zhang:2017:spie}
explore the directional biases of terrestrial lasers in more detail,
in the context of historical cometary orbits.

\begin{figure}
\begin{centering}
\plotone{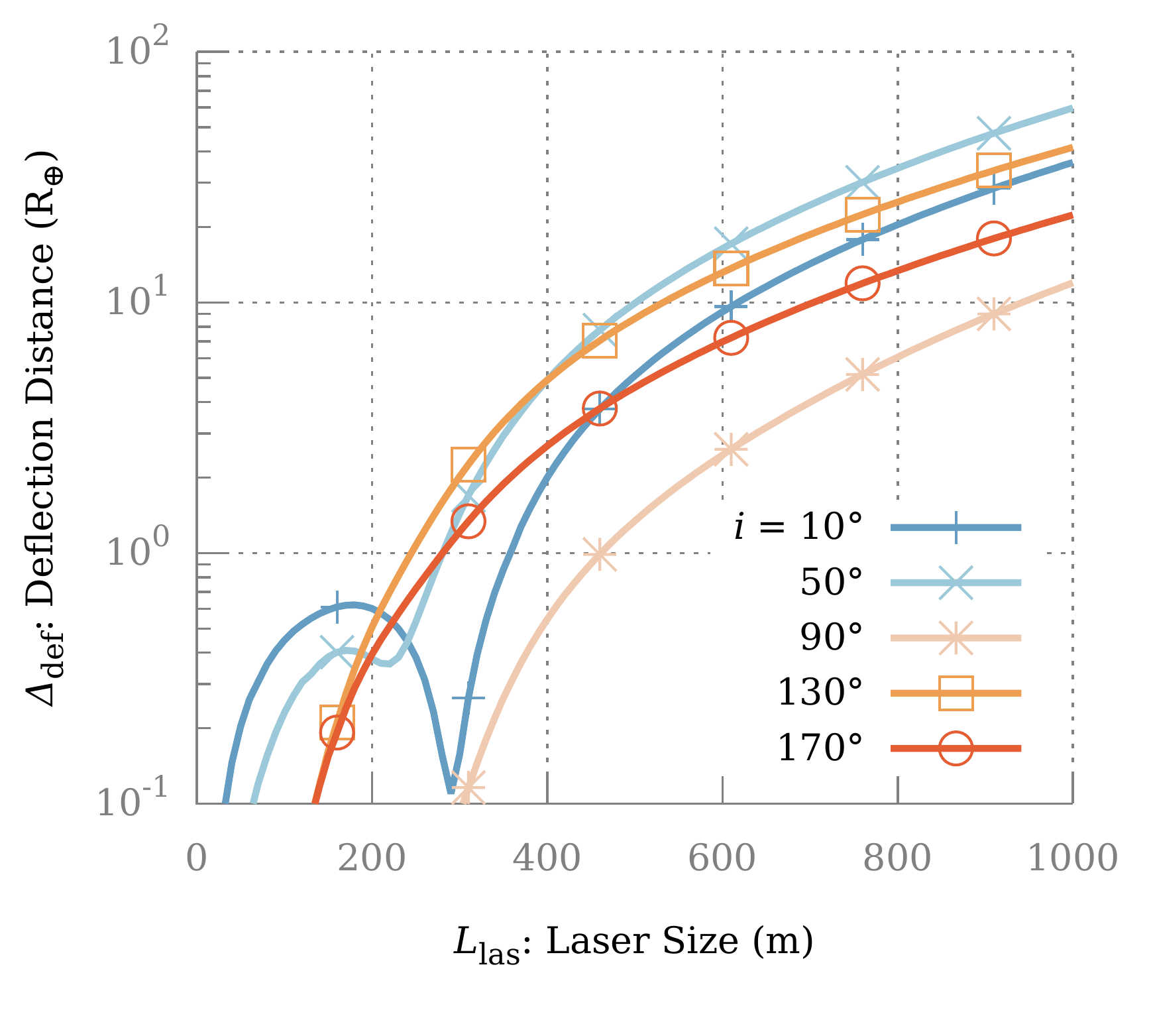}\caption{Deflection effectiveness of $P=1$~GW terrestrial laser arrays of
a range of sizes located at $\phi_{\text{las}}=30\degree\text{S}$
with $\Theta_{\text{fov}}=90\degree$ for comets of various inclinations
$i$ over 2~yr: At $L_{\text{las}}\lesssim200$~m, beam divergence
prevents significant deflection until a few months prior to encounter,
a period where deflection opposes earlier deflection as shown in Fig.~\ref{fig:orbit-branches},
resulting in the interval of negative slope near $L_{\text{las}}\sim200$~m
for the prograde comets. \label{fig:ground-size}}
\par\end{centering}
\end{figure}

Increasing $L_{\text{las}}$, which provides a tighter beam, will
also improve deflection, even without a corresponding increase in
$P$. Fig.~\ref{fig:ground-size} compares the deflection effectiveness
by $P=1$~GW arrays over a range of $L_{\text{las}}$. Increasing
from $L_{\text{las}}=500$~m to $L_{\text{las}}=1$~km boosts the
deflection to a very safe $\Delta_{\text{def}}=10$\textendash 30~$R_{\oplus}$,
depending on the comet's orbit.

\begin{figure}
\begin{centering}
\plotone{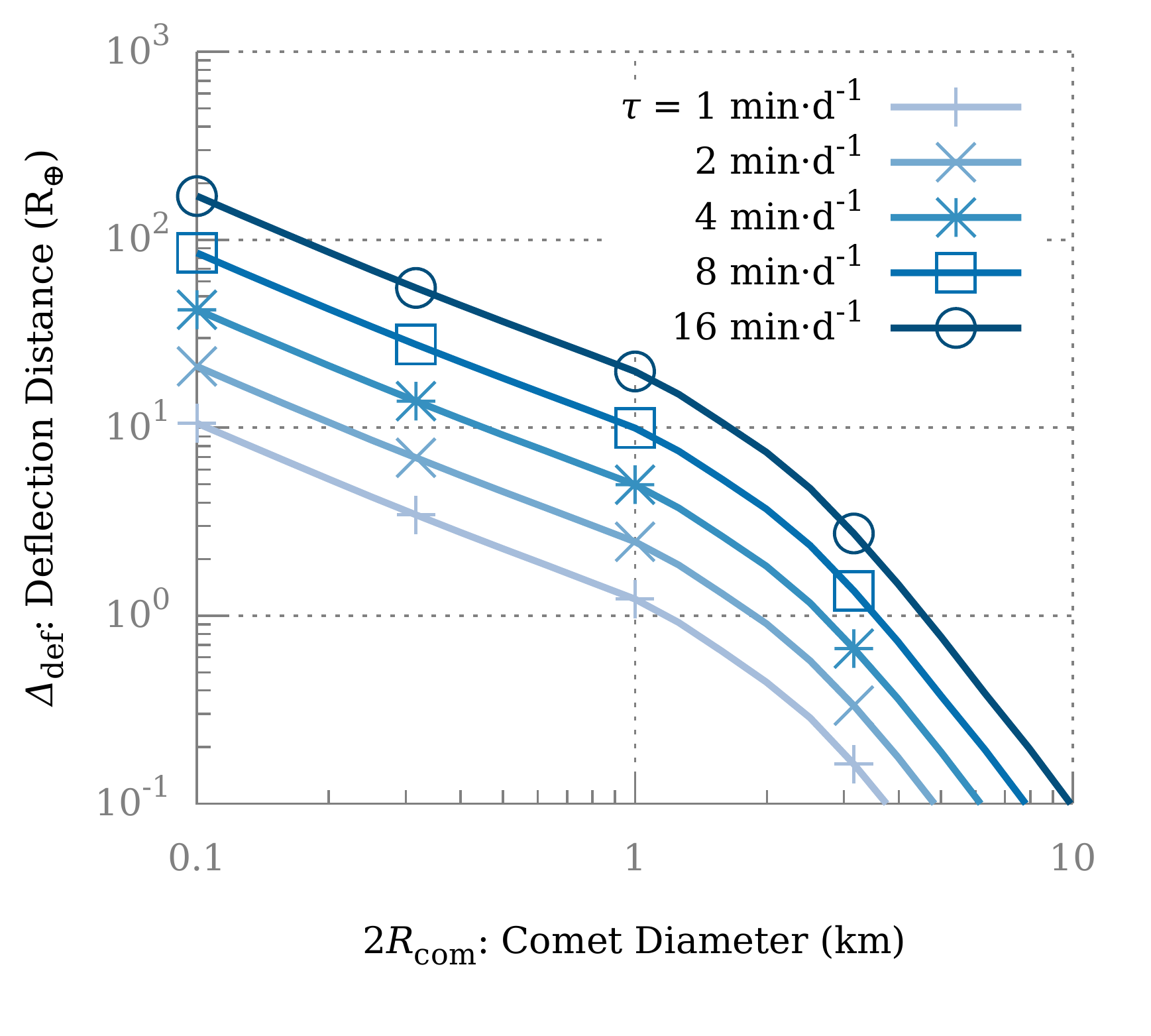}
\par\end{centering}
\caption{A single $L_{\text{las}}=1$~km operating at $P=100$~GW for a few
minutes each day over 2~yr is sufficient to deflect many comets.
At $\tau=4$~min~d$^{-1}$, such a system can deflect the canonical
$2R_{\text{com}}=500$~m comet by a comfortable $\Delta_{\text{def}}\sim10$~$R_{\oplus}$.
Targeting of the comet is assumed to be possible for $\protect\geq\tau$
throughout the deflection period. \label{fig:ground-large}}

\end{figure}

Extremely large and powerful ground-based laser arrays of $L_{\text{las}}=1$~km
and $P=100$~GW have been proposed to enable near-relativistic spaceflight
by radiation pressure on thin, reflective sails \citep{lubin:2016,kulkarni2018}.
Such laser arrays, however, may only operate for a short fraction
$\tau\ll1$ of each day ($P/\langle P\rangle=\tau^{-1}\gg1$). Fig.~\ref{fig:ground-large}
compares deflection for $\tau=1$~min~d$^{-1}$ to 16~min~d$^{-1}$.
An array at $\tau=2$~min~d$^{-1}=1/720$, installed at an appropriate
site, can safely deflect a canonical $2R_{\text{com}}=500$~m comet
by a comfortable $\Delta_{\text{def}}\sim10$~$R_{\oplus}$ over
1~yr. With $\tau=16$~min~d$^{-1}=1/90$, the same laser can deflect
a $2R_{\text{com}}=1$~km by approximately the same distance.

Ultimately, regardless of its power, size and location, a single terrestrial
laser array is insufficient as a long-term solution for comet deflection,
due its limited field of view. A robust terrestrial planetary defense
system will require multiple laser arrays distributed across a wide
enough range of latitudes to provide full sky coverage. With such
a network, every point on the celestial sphere is in the field of
view of at least one laser at some point each day, ensuring the ability
to target comets approaching in any orbit.

\subsection{Fragmentation Mitigation}

Active comets\textemdash especially dynamically new comets entering
the inner solar system for the first time\textemdash have a propensity
to disintegrate under solar heating \citep{boehnhardt:2004}. Laser
heating of the nucleus supplements the natural solar heating, elevating
the likelihood of fragmentation. When presented with a threatening
comet with insufficient notice to carry out a successful deflection,
laser-induced fragmentation could be used, if the consequence of impact
by an intact nucleus is determined to be more severe than impact by
multiple small fragments. Fragmentation, however, hinders a clean
deflection\textemdash the focus of these simulations\textemdash converting
a single nucleus into several smaller nuclei that must be deflected
simultaneously, which may not be possible.

Beyond a few specific instances, the process of comet splitting is
not well understood. Circumstances for disintegration vary dramatically
between comets, with some surviving until within a few radii of the
Sun, and others fragmenting well beyond the orbit of the Earth \citep{boehnhardt:2004}.
The mechanisms proposed for fragmentation under illumination generally
reflect the following pattern:
\begin{enumerate}
\item Cumulative loss of volatiles from the nucleus, weakening its structure;
\item Stress from sublimation pressure overcoming the remaining strength
of the nucleus.
\end{enumerate}
These simulations treat the comet as time-independent, with fixed
$A$ throughout the entire deflection process. This assumption is
valid when volatile loss during deflection\textemdash comparable to,
at most, the volatile loss expected during one perihelion passage
for the scenarios considered\textemdash is negligible compared to
the total mass of volatiles available for sublimation in the nucleus.
Under this condition, the strength of the nucleus remains nearly constant
throughout deflection, and fragmentation can be avoided by limiting
the stress exerted on the nucleus.

\begin{figure}
\begin{centering}
\plotone{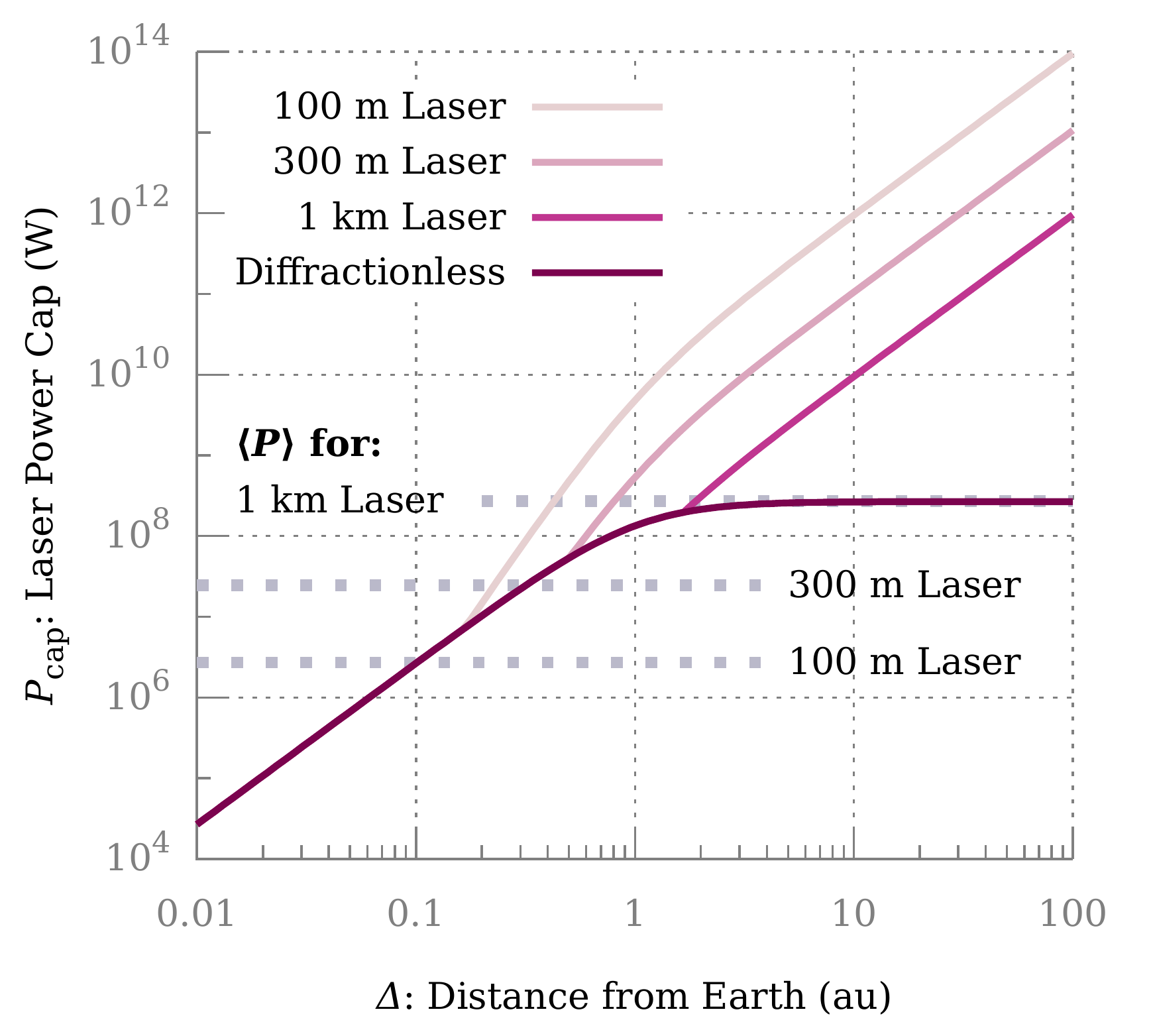}
\par\end{centering}
\caption{A $2R_{\text{com}}=500$~m diameter comet at $r=r_{\text{cap}}=1$~au
intercepts $\sim$3~kW solar radiation. The solid lines indicate
the laser power $P=P_{\text{cap}}$ that, in concert with solar illumination,
produces this maximum heating at various distances $\Delta$ from
the Earth for $L_{\text{las}}=100$~m, 300~m, and 1~km. The ``diffractionless''
curve represents a hypothetical laser with zero beam divergence, where
the entire beam is always intercepted by the comet regardless of $\Delta$.
As shown in the graph, $P_{\text{cap}}$ diverges from the diffractionless
limit once $\Delta$ is sufficiently large for $R_{\text{spot}}>R_{\text{com}}$.
Dotted lines indicate the $\langle P\rangle$ of orbital lasers with
$L_{\text{PV}}=L_{\text{las}}$ and $\varepsilon=20\%$. Distance
from the Sun, used to compute solar flux, is approximated for this
figure as $r=\sqrt{\Delta^{2}+(\text{1 au})^{2}}$. \label{fig:cap}}
\end{figure}

The stress applied to a comet by sublimation is a complicated function
of the comet's geometry and internal structure\textemdash information
unlikely to be available for a newly identified comet. Without this
information, an accurate prediction of disintegration cannot be developed.
When approximating stress as a monotonic function of the total incident
radiation, a straightforward method to avoid disintegration is to
cap the total power incident on the comet to a level such that the
structural integrity of the nucleus is retained. \citet{bortle:1991}
empirically estimated this limit, finding that $\sim$70\% of ground-observed
comets with perihelion distance $q$ and absolute magnitude $H_{0}>7.0+6.0(q/\text{1 au})$
disintegrate. In the absence of a reliable function connecting a comet's
absolute magnitude $H_{0}$ to its radius $R_{\text{com}}$, this
relation cannot be directly incorporated into the simulations. The
relation does, however, suggest that bright (and therefore, large)
comets more readily survive perihelion and are thus more resistant
to fragmentation on heating than their fainter counterparts.

\begin{figure*}
\begin{centering}
\plottwo{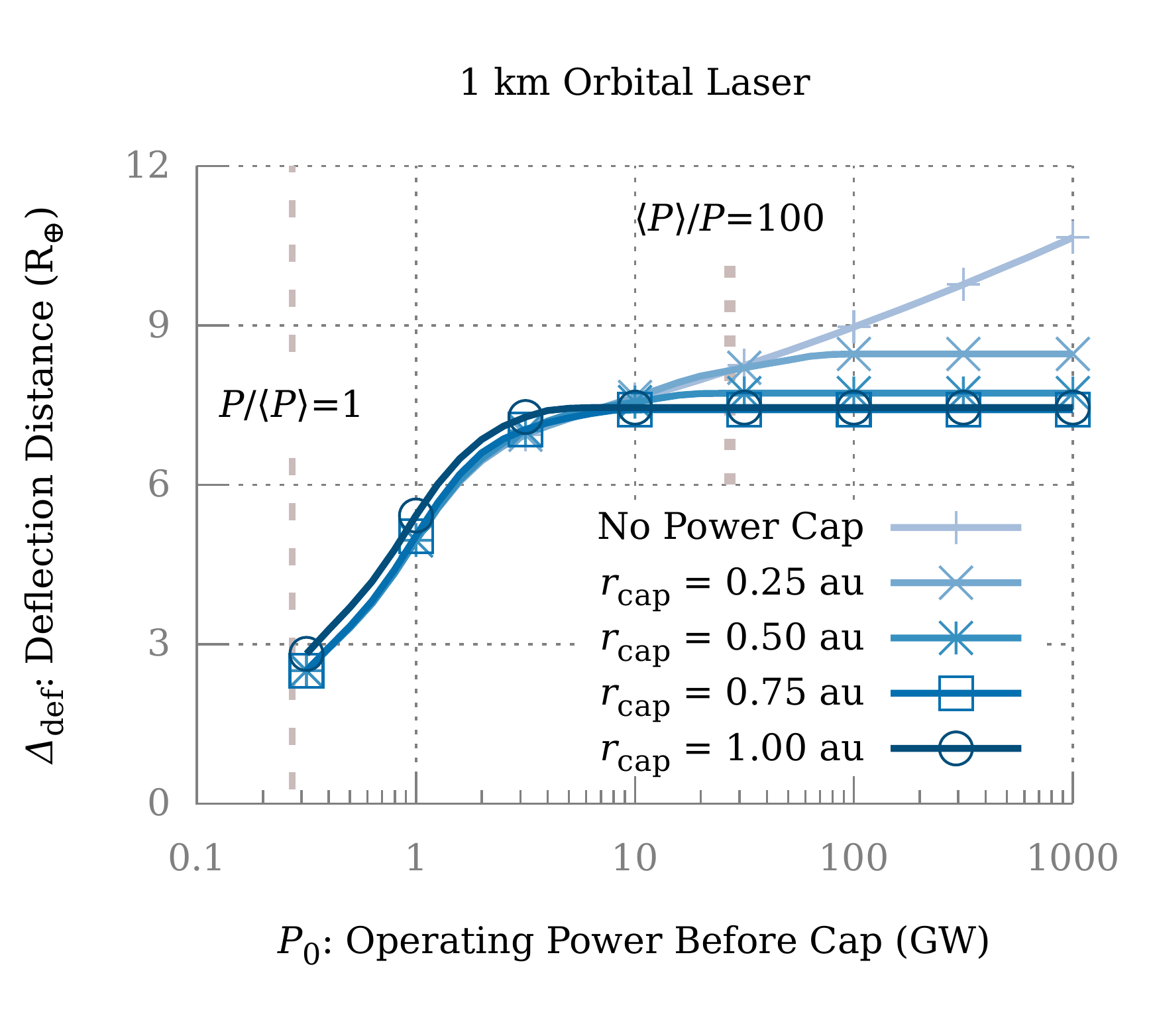}{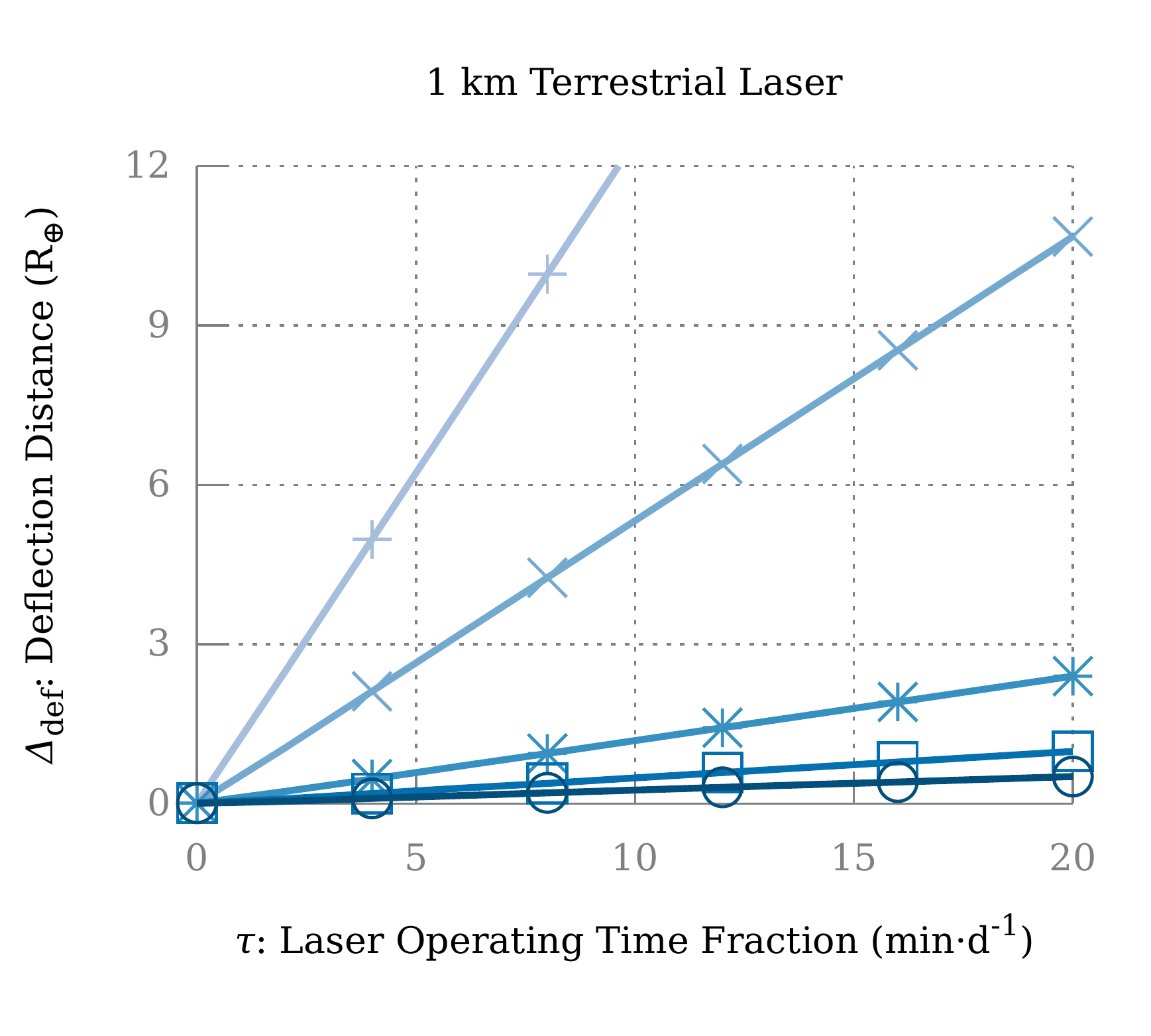}
\par\end{centering}
\caption{Deflection of the canonical comet over 2~yr with a cap on $P$ for
a $L_{\text{las}}=1$~km orbital laser array with $\varepsilon=20\%$
(\textit{left}) and a $L_{\text{las}}=1$~km, $P=1$~GW terrestrial
laser array (\textit{right}): Even a stringent $r_{\text{cap}}=1$~km
only minimally reduces the deflection effectiveness of the orbital
laser, as $\langle P\rangle$ is unaffected. In contrast, terrestrial
lasers are constrained by $\tau$, so a cap on $P$ also sets a cap
on $\langle P\rangle$ which results in a much greater reduction in
effectiveness, particularly for large laser arrays. \label{fig:cap-results}}
\end{figure*}

One strategy could be to restrict laser power to $P\leq P_{\text{cap}}$
where $P=P_{\text{cap}}$ yields a total incident radiation on the
comet, from the Sun and laser combined, equivalent to that from the
Sun alone at $r=r_{\text{cap}}=1$~au, the largest (and hence, safest)
sensible $r_{\text{cap}}$ for avoiding fragmentation during deflection.
Any comet that disintegrates at a larger $r>1$~au would disintegrate
before reaching Earth, even without the laser. Fig.~\ref{fig:cap}
shows $P_{\text{cap}}$ for a $2R_{\text{com}}=500$~m diameter comet
with this $r_{\text{cap}}=1$~au over a range of $\Delta$ for several
$L_{\text{las}}$. If impact is set to occur after the comet's perihelion
passage, or if the comet is known to have previously survived perihelion
at a distance $q$ from the Sun, a less restrictive $r_{\text{cap}}=q\leq1$~au
may be used instead. Meanwhile, if the comet is very bright ($H_{0}\ll7.0$),
the risk of fragmentation from heating, as determined by \citet{bortle:1991},
is sufficient low that a cap on power is unnecessary.

The effects of introducing a $P\leq P_{\text{cap}}$ cap are illustrated
in Fig.~\ref{fig:cap-results}. Such a cap only minimally reduces
the effectiveness of a space laser with fixed $\langle P\rangle$.
In contrast, ground lasers are constrained by $\tau$, and with fixed
$\tau$, a cap on $P$ also sets a cap on $\langle P\rangle$ which
produces a much larger reduction in deflection effectiveness. This
effect is especially pronounced for large laser arrays, for where
the cap is active for comets well beyond the inner solar system. Beginning
deflection early\textemdash while the comet is still sufficiently
far for the cap to be inactive\textemdash overcomes this limitation,
although doing so would require significant advancement in detection
and tracking capability. Alternatively, if fragmentation can be predicted
more reliably, a less stringent cap may be used.

Note that, given a sufficiently early discovery and sufficiently large
laser, it may still be possible to deflect a comet with an even higher
$r_{\text{cap}}>1$~au, such that its trajectory has shifted sufficiently
to avoid impact by all fragments before it reaches $r<r_{\text{cap}}$
and fragments from solar heating. Simulating this scenario, however,
requires modeling the fragmentation process of a comet nucleus, a
process better suited for a more detailed analysis of the general
considerations for planetary defense from cometary impacts.

Finally, it is important to recognize that, although these considerations
may reduce the likelihood of fragmentation, they will not fully eliminate
the risk. The residual risk poses a significant challenge that must
eventually be addressed prior to the commencement of deflection. Note
also that, depending on the specifics of the threat, fragmentation
of the comet nucleus may be preferable to the impact of an intact
nucleus. Intentional disruption may be achievable by following the
opposite of the strategies above, elevating $P$ to point where the
tensile strength of the nucleus is exceeded. Both topics require separate
in-depth analyses and will be deferred to future analyses of planetary
defense strategy for cometary impactors.

\section{Conclusions}

Comets pose unique challenges left unanswered by most techniques for
mitigating asteroid impact. Comets' highly eccentric orbits hinder
discovery until, at best, a few years before impact. The expected
uncertainties in trajectory for a newly discovered object, including
in $A$, introduce further delays to a response. The rapid progression
from discovery to impact, coupled with often extreme delta-v requirements,
renders interception missions either unreliable or otherwise impractical
with presently available propulsion technologies.

This lack of attention stems in part from the rarity of comets relative
to NEAs. Comets of all groups are estimated to be responsible for
less than 1\% of all impact events in Earth's recent geological record,
though they may comprise the majority of large impactors of diameter
$2R_{\text{com}}\gtrsim1$~km \citep{yeomans:2013}. No comets of
any size have been confirmed to have impacted the Earth in the historical
past, nor is one expected to impact anytime in the foreseeable future.
Hence, the near-term risk posed by comets is far lower than that of
asteroids, which are generally smaller but far more common\textemdash and
have been observed to impact the Earth in recent history.

Even so, given their unpredictable timing and the likely catastrophic
global consequences of an impact, comet deflection remains an important
consideration in planetary defense strategy. Further attention should
be given to the possibility and consequences of comet disintegration
during deflection\textemdash as well as other unintended consequences,
such as dust generation, that may prove fatal to insufficiently shielded
satellites in Earth orbit \citep{beech:1995}.

Attention must also be directed toward the engineering challenges
of large-scale laser arrays. Unless a strategy is prepared and a system
is developed and primed before discovery, impact mitigation will be
improbable. However, given adequate preparation, these preliminary
simulations suggest that use of large, high-powered laser arrays\textemdash either
in Earth orbit or, with advancements in adaptive optics technology,
on the ground\textemdash may prove to be a viable strategy for mitigating
comet impacts.

\acknowledgements{We gratefully acknowledge funding from NASA California Space Grant
grant NNX10AT93H and from NASA Innovative Advanced Concepts grants
NNX15AL91G and NNX16AL32G as well as a generous gift from the Emmett
and Gladys W. Fund in support of this research. We also thank the
anonymous referee whose detailed comments and suggestions helped improve
this manuscript. An earlier version of some of these results was presented
at SPIE Optics + Photonics in August 2016 in San Diego, CA \citep{zhang:2016:spie}.}

\software{GNU Parallel \citep{tange:2011}, gnuplot \citep{williams:2017}}

\bibliographystyle{aasjournal}
\bibliography{ref}

\end{document}